\documentclass[12pt]{article}
\usepackage[numbers,sort&compress]{natbib}
\usepackage[numbers]{pstricks}
\usepackage{placeins}
\usepackage{epsfig}
\usepackage{epstopdf}
\usepackage{amsmath}
\usepackage{hhline}
\usepackage{amssymb}
\usepackage{times}
\usepackage{array}
\usepackage{rotating}
\newcommand{\includegraph}[2][]{\includegraphics[#1]{#2}}

\newlength{\dinwidth}
\newlength{\dinmargin}
\begin{document}  
\newcommand{\pom}{{I\!\!P}}
\newcommand{\reg}{{I\!\!R}}
\newcommand{\slowpi}{\pi_{\mathit{slow}}}
\newcommand{\fiidiii}{F_2^{D(3)}}
\newcommand{\fiidiiiarg}{\fiidiii\,(\beta,\,Q^2,\,x)}
\newcommand{\n}{1.19\pm 0.06 (stat.) \pm0.07 (syst.)}
\newcommand{\nz}{1.30\pm 0.08 (stat.)^{+0.08}_{-0.14} (syst.)}
\newcommand{\fiidiiiful}{F_2^{D(4)}\,(\beta,\,Q^2,\,x,\,t)}
\newcommand{\fiipom}{\tilde F_2^D}
\newcommand{\ALPHA}{1.10\pm0.03 (stat.) \pm0.04 (syst.)}
\newcommand{\ALPHAZ}{1.15\pm0.04 (stat.)^{+0.04}_{-0.07} (syst.)}
\newcommand{\fiipomarg}{\fiipom\,(\beta,\,Q^2)}
\newcommand{\pomflux}{f_{\pom / p}}
\newcommand{\nxpom}{1.19\pm 0.06 (stat.) \pm0.07 (syst.)}
\newcommand {\gapprox}
   {\raisebox{-0.7ex}{$\stackrel {\textstyle>}{\sim}$}}
\newcommand {\lapprox}
   {\raisebox{-0.7ex}{$\stackrel {\textstyle<}{\sim}$}}
\def\gsim{\,\lower.25ex\hbox{$\scriptstyle\sim$}\kern-1.30ex%
\raise 0.55ex\hbox{$\scriptstyle >$}\,}
\def\lsim{\,\lower.25ex\hbox{$\scriptstyle\sim$}\kern-1.30ex%
\raise 0.55ex\hbox{$\scriptstyle <$}\,}
\newcommand{\pomfluxarg}{f_{\pom / p}\,(x_\pom)}
\newcommand{\dsf}{\mbox{$F_2^{D(3)}$}}
\newcommand{\dsfva}{\mbox{$F_2^{D(3)}(\beta,Q^2,x_{I\!\!P})$}}
\newcommand{\dsfvb}{\mbox{$F_2^{D(3)}(\beta,Q^2,x)$}}
\newcommand{\dsfpom}{$F_2^{I\!\!P}$}
\newcommand{\gap}{\stackrel{>}{\sim}}
\newcommand{\lap}{\stackrel{<}{\sim}}
\newcommand{\fem}{$F_2^{em}$}
\newcommand{\tsnmp}{$\tilde{\sigma}_{NC}(e^{\mp})$}
\newcommand{\tsnm}{$\tilde{\sigma}_{NC}(e^-)$}
\newcommand{\tsnp}{$\tilde{\sigma}_{NC}(e^+)$}
\newcommand{\st}{$\star$}
\newcommand{\sst}{$\star \star$}
\newcommand{\ssst}{$\star \star \star$}
\newcommand{\sssst}{$\star \star \star \star$}
\newcommand{\tw}{\theta_W}
\newcommand{\sw}{\sin{\theta_W}}
\newcommand{\cw}{\cos{\theta_W}}
\newcommand{\sww}{\sin^2{\theta_W}}
\newcommand{\cww}{\cos^2{\theta_W}}
\newcommand{\trm}{m_{\perp}}
\newcommand{\trp}{p_{\perp}}
\newcommand{\trmm}{m_{\perp}^2}
\newcommand{\trpp}{p_{\perp}^2}
\newcommand{\alp}{\alpha_s}

\newcommand{\alps}{\alpha_s}
\newcommand{\sqrts}{$\sqrt{s}$}
\newcommand{\LO}{$O(\alpha_s^0)$}
\newcommand{\Oa}{$O(\alpha_s)$}
\newcommand{\Oaa}{$O(\alpha_s^2)$}
\newcommand{\PT}{p_{\perp}}
\newcommand{\JPSI}{J/\psi}
\newcommand{\sh}{\hat{s}}
\newcommand{\uh}{\hat{u}}
\newcommand{\MP}{m_{J/\psi}}
\newcommand{\PO}{I\!\!P}
\newcommand{\xbj}{x}
\newcommand{\xpom}{x_{\PO}}
\newcommand{\ttbs}{\char'134}
\newcommand{\xpomlo}{3\times10^{-4}}  
\newcommand{\xpomup}{0.05}  
\newcommand{\dgr}{^\circ}
\newcommand{\pbarnt}{\,\mbox{{\rm pb$^{-1}$}}}
\newcommand{\gev}{\,\mbox{GeV}}
\newcommand{\WBoson}{\mbox{$W$}}
\newcommand{\fbarn}{\,\mbox{{\rm fb}}}
\newcommand{\fbarnt}{\,\mbox{{\rm fb$^{-1}$}}}
\newcommand{\dsdx}[1]{$d\sigma\!/\!d #1\,$}
\newcommand{\eV}{\mbox{e\hspace{-0.08em}V}}
%
%
\newcommand{\qsq}{\ensuremath{Q^2} }
\newcommand{\gevsq}{\ensuremath{\mathrm{GeV}^2} }
\newcommand{\et}{\ensuremath{E_t^*} }
\newcommand{\rap}{\ensuremath{\eta^*} }
\newcommand{\gp}{\ensuremath{\gamma^*}p }
\newcommand{\dsiget}{\ensuremath{{\rm d}\sigma_{ep}/{\rm d}E_t^*} }
\newcommand{\dsigrap}{\ensuremath{{\rm d}\sigma_{ep}/{\rm d}\eta^*} }

\newcommand{\dstar}{\ensuremath{D^*}}
\newcommand{\dstarp}{\ensuremath{D^{*+}}}
\newcommand{\dstarm}{\ensuremath{D^{*-}}}
\newcommand{\dstarpm}{\ensuremath{D^{*\pm}}}
\newcommand{\zDs}{\ensuremath{z(\dstar )}}
\newcommand{\Wgp}{\ensuremath{W_{\gamma p}}}
\newcommand{\ptds}{\ensuremath{p_t(\dstar )}}
\newcommand{\etads}{\ensuremath{\eta(\dstar )}}
\newcommand{\ptj}{\ensuremath{p_t(\mbox{jet})}}
\newcommand{\ptjn}[1]{\ensuremath{p_t(\mbox{jet$_{#1}$})}}
\newcommand{\etaj}{\ensuremath{\eta(\mbox{jet})}}
\newcommand{\detadsj}{\ensuremath{\eta(\dstar )\, \mbox{-}\, \etaj}}

\newcommand{\TODO}[1]{{\\ {\red #1}\\ }}

\def\Journal#1#2#3#4{{#1} {\bf #2} (#3) #4}
\def\NCA{\em Nuovo Cimento}
\def\NIM{\em Nucl. Instrum. Methods}
\def\NIMA{{\em Nucl. Instrum. Methods} {\bf A}}
\def\NPB{{\em Nucl. Phys.}   {\bf B}}
\def\PLB{{\em Phys. Lett.}   {\bf B}}
\def\PRL{\em Phys. Rev. Lett.}
\def\PRD{{\em Phys. Rev.}    {\bf D}}
\def\ZPC{{\em Z. Phys.}      {\bf C}}
\def\EJC{{\em Eur. Phys. J.} {\bf C}}
\def\CPC{\em Comp. Phys. Commun.}

\begin{titlepage}

\noindent
\begin{flushleft}
{\tt DESY 14-242    \hfill    ISSN 0418-9833} \\
{\tt December 2014}                  \\
\end{flushleft}

\noindent
\\
\noindent

\vspace{2cm}
\begin{center}
\begin{Large}

{\bf Diffractive Dijet Production with a Leading Proton in \begin{boldmath}$ep$\end{boldmath} Collisions at HERA}

\vspace{2cm}

H1 Collaboration

\end{Large}
\end{center}

\vspace{2cm}

\begin{abstract}
  The cross section of the diffractive process $e^{+}p \to
  e^{+}Xp$ is measured at a centre-of-mass energy of
  $318\,\text{GeV}$, where the system $X$ contains at least two
  jets and the leading final state proton $p$ is detected in the
  H1 Very Forward Proton Spectrometer.
  The measurement is performed in photoproduction with photon
  virtualities $Q^2 <2\,\text{GeV}^2$ and in deep-inelastic scattering
  with $4\,\text{GeV}^2<Q^2<80\,\text{GeV}^2$. The results are
  compared to next-to-leading order QCD calculations based on
  diffractive parton distribution functions as extracted from
  measurements of inclusive cross sections in diffractive
  deep-inelastic scattering.
\end{abstract}

\vspace{1.5cm}

\begin{center}
Submitted to {\it JHEP} 
\end{center}

\end{titlepage}

%
%
%
\begin{flushleft}

V.~Andreev$^{21}$,             
A.~Baghdasaryan$^{33}$,        
K.~Begzsuren$^{30}$,           
A.~Belousov$^{21}$,            
P.~Belov$^{10}$,               
V.~Boudry$^{24}$,              
G.~Brandt$^{45}$,              
M.~Brinkmann$^{10}$,           
V.~Brisson$^{23}$,             
D.~Britzger$^{10}$
% ...


\bigskip{\it
% ...
$ ^{10}$ DESY, Hamburg, Germany \\
% ...
$ ^{21}$ Lebedev Physical Institute, Moscow, Russia \\
% ...
$ ^{23}$ LAL, Universit\'e Paris-Sud, CNRS/IN2P3, Orsay, France \\
 $ ^{24}$ LLR, Ecole Polytechnique, CNRS/IN2P3, Palaiseau, France \\
% ...
$ ^{30}$ Institute of Physics and Technology of the Mongolian
          Academy of Sciences, Ulaanbaatar, Mongolia \\
% ...
$ ^{33}$ Yerevan Physics Institute, Yerevan, Armenia \\
% ...
$ ^{45}$ Department of Physics, Oxford University,
          Oxford, UK$^{ b}$ \\

\bigskip
% ...
$ ^b$ Supported by the UK Science and Technology Facilities Council,
      and formerly by the UK Particle Physics and
      Astronomy Research Council \\
% ...
}\end{flushleft}
%

\newpage

\section{Introduction}
Diffractive processes, $ep \to eXY$, where the systems $X$ and $Y$ are
separated in rapidity, have been studied extensively at the
electron-proton collider HERA.
In diffractive processes the interacting hadrons remain intact or
dissociate into low mass hadronic systems via an exchange which has
vacuum quantum numbers, often referred to as a pomeron ($\pom$).
 Experimentally, 
 diffractive events may be selected 
 either by the presence of a large rapidity gap (LRG) in the rapidity distribution of the
 outgoing hadrons or by detecting a leading proton in the final state.
The H1 experiment was equipped with two dedicated detectors, the
Forward Proton Spectrometer (FPS) \cite{VanEsch:1999pi} and the Very Forward
Proton Spectrometer (VFPS) \cite{Astvatsatourov:2014dna} to detect the
leading protons.

 In the framework of the collinear factorisation theorem
 \cite{Collins:1997sr}  diffractive parton distribution functions
 (DPDFs) may be defined.
 The factorisation theorem predicts that the cross section can be
 expressed as the convolution of non-perturbative DPDFs and partonic
 cross sections of the hard sub-process, calculable within perturbative
 Quantum Chromodynamics (QCD).
 The DPDFs have properties similar to the parton distribution
 functions of the proton, but with the constraint of a leading proton
 or its low mass excitations being present in the final state.

DPDFs were obtained at HERA from inclusive diffractive deep-inelastic
scattering (DDIS) data \cite{Aktas:2006hy,Chekanov:2004hy}.
Given the DPDFs, perturbative QCD calculations are expected to be
applicable to other processes such as jet and heavy quark production
in DDIS at HERA
\cite{Chekanov:2002qm,Aktas:2006up,Aktas:2007bv,Chekanov:2007aa,Aaron:2011mp,boris}.
Indeed, next-to-leading order (NLO) QCD predictions using DPDFs
describe these measurements well.

In diffractive hadron-hadron interactions however, the production of
jets is found to be suppressed by about one order of magnitude
\cite{Affolder:2000vb,Chatrchyan:2012vc}, as compared to predictions
based on HERA DPDFs.
This "factorisation breaking" may be explained e.g.\ by soft interactions or multi-pomeron
exchanges between the hadrons and/or rescattering phenomena which destroy the diffractive event
signature \cite{Kaidalov:2001iz,Kaidalov:2003gy,Collins:2001ga}.

The issues of DPDF applicability and factorisation breaking can also
be studied in hard diffractive photoproduction ($\gamma p$), where
the virtuality of the exchanged photon $Q^2$  is close to zero.
In the photoproduction regime, within the leading order approach, the
small photon virtuality allows for partonic fluctuations $\gamma  \to
q\overline{q}$ that last long enough to interact with the partons in
the proton. In this regime the photon can be treated as a quasi-real target
and therefore exhibits hadronic structure.

Diffractive photoproduction of dijets in $ep$ collisions at HERA have
been measured by H1 \cite{Aaron:2010su,Aktas:2007hn} and ZEUS
\cite{Chekanov:2007rh}.
In each of these measurements diffractive events are selected by
requiring a large rapidity gap.
Different ratios of data to the NLO QCD prediction have been reported
by H1 and ZEUS: while H1
reported their data to be suppressed by a factor of $0.6$ with respect
to the NLO QCD predictions \cite{Aaron:2010su,Aktas:2007hn}, the ZEUS data
are compatible with the theoretical expectations 
\cite{Chekanov:2009aa}.
%
Various mechanisms of suppressing diffractive dijet 
photoproduction have been proposed \cite{Kaidalov:2003xf,Kaidalov:2009fp}.

Enhanced sensitivity to the differences between theory and data may be
achieved by calculating the double ratio of the ratio of data to predictions of diffractive dijet photoproduction to the corresponding ratio in DDIS \cite{Aktas:2007hn}.
In this way several experimental systematic uncertainties cancel and theoretical uncertainties
can be reduced.

In the present paper new measurements of diffractive dijet cross
sections in DIS and photoproduction are presented.
The data were collected in the years $2006$ and $2007$ with a total integrated
luminosity of $30\,\text{pb}^{-1}$ for diffractive photoproduction and
$50\,\text{pb}^{-1}$ for diffractive DIS.
For the identification of diffractive events a proton detected in the
VFPS is required.
The results are compared to NLO QCD calculations.

\section{Kinematics}
\begin{figure}
\center
\includegraph[width=0.9\textwidth]{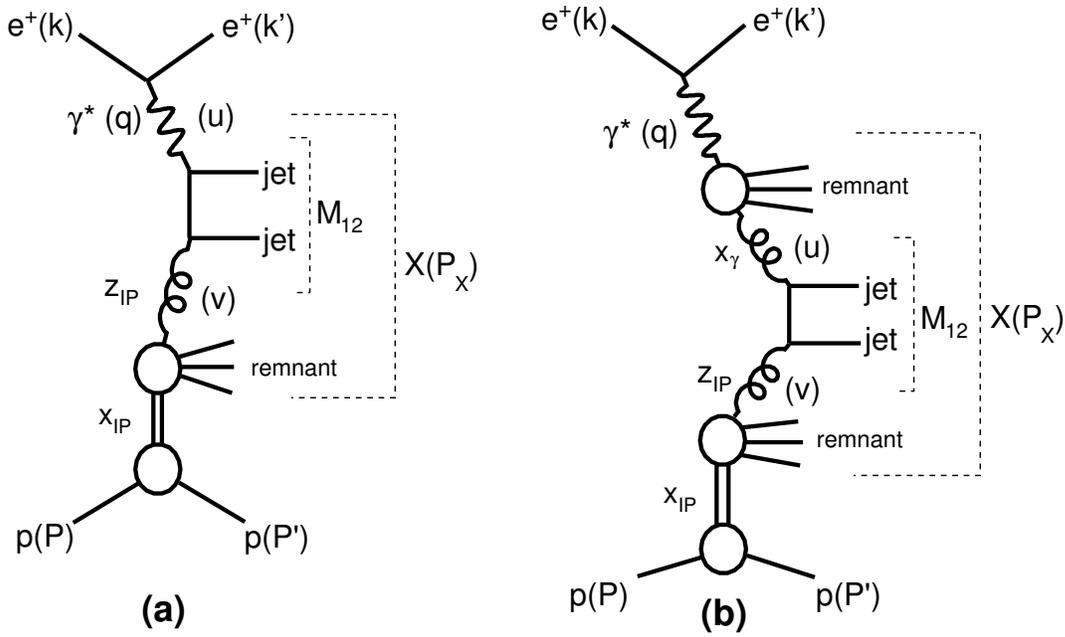} 
\caption{Leading order diagrams of the direct a) and resolved b)
  diffractive dijet production.}
\label{fig_feyn_dir_res}
\end{figure}

Figures~\ref{fig_feyn_dir_res} (a) and (b) show leading order
diagrams of direct and resolved diffractive dijet production in
$ep$ interactions.
The relative contributions of these two components depend on the virtuality
of the exchanged photon such that at high virtualities the direct process
is dominating.
The incoming (scattered) positron four-momentum is denoted as $k$
($k'$), the four-momentum of the virtual photon emitted from the
positron as $q = k - k'$.
The four-momentum of the incoming (outgoing) proton is $P$ ($P'$).
The kinematics of the $ep$ scattering process can be described
by
\begin{equation}
s = (k+P)^2, \qquad Q^2= - q^2,   \qquad y = \frac{P\cdot q}{P\cdot k}~,
\label{eqn_dis}
\end{equation}
where $s$ is the square of the centre-of-mass energy of the collision,
$Q^2$ is the photon virtuality and $y$ the inelasticity of the
process.
With $P_X$ being  the four-momentum of the hadronic final state
excluding the leading proton (see figure \ref{fig_feyn_dir_res}), the
inclusive diffractive kinematics is described by the additional
variables
\begin{equation}
M_X^2 = P_X^2, \qquad x_{\pom} = \frac{q \cdot (P-P')}{q \cdot P},\qquad t = (P - P')^2,
\end{equation}
where $M_X$ is the invariant mass of system $X$, $x_{\pom}$
corresponds to the longitudinal momentum fraction lost by the incoming
proton and  $t$ is the four-momentum transfer squared at the proton
vertex.

For diffractive dijet production additional invariants are introduced.
With denoting the four-momenta entering the hard sub-process from the
photon and from the pomeron side as $u$ and  $v$, the longitudinal fractions of the
photon and of the pomeron momentum entering the hard sub-process, $x_{\gamma}$ and
$z_{\pom}$, are defined as
\begin{equation}
x_{\gamma} = \frac{P\cdot u}{P \cdot q} \qquad{\rm and}\qquad z_{\pom} = \frac{q\cdot v}{q \cdot (P-P')},
\end{equation}
respectively.

In leading order, the invariant mass of the dijet system $M_{12}$ is
equal to the centre-of-mass energy of the hard sub-process
\begin{equation}
M_{12}^2 = (u+v)^2.
\end{equation}

\section{Factorisation in Diffractive Dijet Production}
In the QCD factorisation approach the diffractive dijet cross section
is given by the convolution of partonic cross sections
$\mathrm{d}\hat{\sigma}$ with diffractive parton distributions
$f_{i/p}^D$:
\begin{eqnarray}
\mathrm{d}\sigma (ep \to e + 2\, \mathrm{jets} + X' + p) = \sum_i \int \mathrm{d}t \int \mathrm{d}x_{\pom} \int \mathrm{d} z_{\pom} \nonumber \\
 \mathrm{d}\hat{\sigma}_{e i \to 2\, \mathrm{jets}}  (\hat{s},\mu_R^2, \mu_F^2) \times f_{i/p}^D (z_{\pom}, \mu_F^2, x_{\pom}, t).
\end{eqnarray}
Here, the hadronic system $X'$ corresponds to what remains of the
system $X$ after removing the two jets.
The integrals extend over the accepted phase space.
The sum runs over all partons $i$ contributing to the cross section,
$\hat{s}\sim x_{\pom} z_{\pom} y s - Q^2$ is the
sub-process invariant energy squared and $\mu_F$ and $\mu_R$ denote the
factorisation and renormalisation scales, respectively.

In the photoproduction region the exchanged photon may dissociate into
a low mass non-perturbative hadronic system due to its low virtuality
(figure~\ref{fig_feyn_dir_res}b) and a photon parton distribution
function ($\gamma$PDF) is introduced.
The cross section for this resolved photon process is given by
\begin{eqnarray}
\mathrm{d}\sigma (ep \to e + 2\, \mathrm{jets} + X' + p) = \sum_{i,j} \int \mathrm{d}t \int \mathrm{d}x_{\pom} \int \mathrm{d} z_{\pom} \nonumber \\
  \int \mathrm{d}y\; f_{\gamma/e} (y)  \int \mathrm{d}x_{\gamma}\; f_{j/\gamma} (x_{\gamma}, \mu_F^2) \times  \mathrm{d}\hat{\sigma}_{i j \to 2\, \mathrm{jets}}  (\hat{s},\mu_R^2, \mu_F^2) \times f_{i/p}^D (z_{\pom}, \mu_F^2, x_{\pom}, t),
\end{eqnarray}
where $f_{\gamma/e}$ is the Weizs\"{a}cker-Williams equivalent photon
flux \cite{vonWeizsacker:1934sx,Williams:1934ad} integrated over
the measured $Q^2$ range and $f_{j/\gamma}$ are the parton
distribution functions in the photon ($\gamma$-PDF). In this case, the centre-of-mass energy
of the hard subprocess is approximated by
$\hat{s}\sim x_{\gamma} x_{\pom} z_{\pom} y s$.
As default, the GRV \cite{Gluck:1991ee} $\gamma$-PDFs
are used to describe the structure of resolved photons.
The AFG \cite{Aurenche:2005da}  $\gamma$-PDF set is also studied.

For the diffractive proton parton densities, the H12006 Fit-B DPDF set 
\cite{Aktas:2006hy} is used.
This parametrisation was obtained from a QCD fit in NLO accuracy to inclusive DDIS data.
In this fit a proton vertex factorisation \cite{Ingelman:1984ns} is assumed in which the 
$x_{\pom}$ and $t$
dependencies of the DPDFs factorise from the dependencies on $\mu_F$ and $z_{\pom}$ such that 
\begin{equation}
f_{i/p}^D (z_{\pom},\mu_F^2,x_{\pom},t) = f_{\pom/p}(x_{\pom},t)\, f_{i/\pom} (z_{\pom}, \mu_F^2) + n_{\reg}\, f_{\reg/p}(x_{\pom},t)\, f_{i/\reg} (z_{\pom}, \mu_F^2).
\label{regge_fact}
\end{equation}
The pomeron flux factor $f_{\pom/p}(x_{\pom},t)$ was parametrised in
\cite{Aktas:2006hy} as suggested by Regge models \cite{Regge:1959mz}.
For $x_\pom \gg 0.01$ a small additional contribution from sub-leading
reggeon ($\reg$) exchange described by the second term
in~(\ref{regge_fact}) was taken into account, where $n_{\reg} \sim
10^{-3}$ is the normalisation factor of the reggeon contribution
\cite{Aktas:2006hy}.

\section{NLO QCD Calculations}
\label{nlo-text}
Theoretical calculations of dijet production in next-to-leading order 
were performed in the $\gamma p$ 
regime using the 
the FKS program~\cite{Frixione:1995ms} and in 
DIS using NLOJET++ \cite{Nagy:1998bb,Nagy:2001xb}.
Both programs were adapted \cite{Aktas:2007hn} for hard diffraction.
The NLO calculations for photoproduction are consistent with 
calculations performed by Klasen and Kramer
\cite{Klasen:1996it,Klasen:2008ah,Cerny:2008zz}.
Similarly, the DDIS predictions were checked using the independent
package DISENT NLO \cite{Chyla:2005qh}.

The NLO calculations are performed with the number of flavours fixed
to 5 and the QCD scale parameter set to $\Lambda_5 = 0.228\, \text{GeV}$,
corresponding to a 2-loop $\alpha_S(M_Z)$ 
of $0.118$.
The renormalisation and factorisation scales are set to be equal and
are calculated from the average jet transverse energy
$\langle
E_T^{*\mathrm{jet}}\rangle=(E_T^{*\mathrm{jet1}}+E_T^{*\mathrm{jet2}})/2$
and the momentum transfer $Q^2$ as $\mu_R^2 = \mu_F^2 =
\langle E_T^{*\mathrm{jet}}\rangle^2+Q^2$.
For photoproduction, $Q^2$ is set to zero.
%
The sensitivity of the NLO predictions to the scale choice is
studied by varying the scale up and down by a factor of
two. An alternative definition of the scale $\mu_R^2 = \mu_F^2 =
(E_T^{*\mathrm{jet1}})^2+Q^2/4$, based on the leading jet
transverse energy $E_T^{*\mathrm{jet1}}$, is also studied.

\section{Experimental Procedure}
\subsection{The H1 Detector}
A detailed  description of the H1 detector can be found elsewhere
\cite{Abt:1996hi,Abt:1996xv,Appuhn:1996na}.
Here only the detector components most relevant to the present
analysis are briefly described.
A right-handed coordinate system is employed with the origin at the
nominal $ep$ interaction point and with the positive $z$-axis pointing
in the proton beam direction.
The $x$-axis is pointing along the horizontal direction to the centre
of the HERA ring.
The pseudorapidity $\eta = -\ln \tan \frac{\theta}{2}$ is calculated
using the polar angle $\theta$ measured with respect to the proton
beam direction.

The interaction point is surrounded by the central tracking detector
(CTD), which consists of a set of concentric drift chambers 
supplemented by silicon
detectors \cite{Pitzl:2000wz} located inside the drift chambers.
Charged particle trajectories are bent by a $1.15\,\text{T}$
homogeneous solenoidal magnetic field.
The region in pseudorapidity covered by the CTD is $-2.0<\eta<2.0$ and
the transverse momentum resolution is $\sigma(p_T)/p_T \simeq
0.002\,p_T/\mbox{GeV}\oplus 0.015$.
A multi-wire proportional chamber at inner radii (CIP) is mainly used
for triggering \cite{Becker:2007ms}.
The forward tracking detector supplements the CTD track
reconstructions in the region $7^{\circ}<\theta < 25^{\circ}$.

Scattered positrons 
in the rapidity range $-4<\eta<-1.4$
are measured in a lead / scintillating-fibre
calorimeter, the SpaCal \cite{Appuhn:1996na}, with energy resolution
$7\%/\sqrt{E/\text{GeV}} \oplus 1\%$.

The central and forward tracking detectors are surrounded by a finely
segmented Liquid Argon (LAr) calorimeter \cite{Andrieu:1993kh}
situated inside the solenoidal magnet and covering the pseudorapidity
region $-1.5<\eta<3.4$.
Its resolution was measured in test beams
\cite{Andrieu:1994yn,Andrieu:1993tz} and is  $11\%/\sqrt{E/\text{GeV}}
\oplus 1\%$ and $50\%/\sqrt{E/\text{GeV}} \oplus 2\%$ for
electromagnetic and hadronic showers, respectively.
%
%
The hadronic energy scale is known within $2\%$ for this analysis
\cite{Salek:2010ifa}.
%

The $ep$ luminosity is determined online by measuring  the event rate
of the Bethe-Heitler brems\-strahlung process, $ep \to ep\gamma$,
where the photon is detected in a calorimeter located close
to the beam pipe at $z = -103\,\text{m}$ \cite{Abt:1996hi}.
The overall integrated luminosity normalisation is determined using a
precision measurement of the QED Compton process \cite{Aaron:2012kn}.

\subsection{Very Forward Proton Spectrometer}
The Very Forward Proton Spectrometer (VFPS) consists of two Roman pots
located $218\,\text{m}$ and $222\,\text{m}$ from the interaction point
in the forward direction.
It allows for a measurement of protons with energies between $895$ and
$912\,\text{GeV}$ ($0.008 < x_{\pom} < 0.028$) and with transverse
momenta up to about $0.8\,\text{GeV}$ ($|t| < 0.6\,\text{GeV}^2$)
\cite{Astvatsatourov:2014dna}.

\begin{figure}[ht]
\center
\includegraph[width=0.6\textwidth]{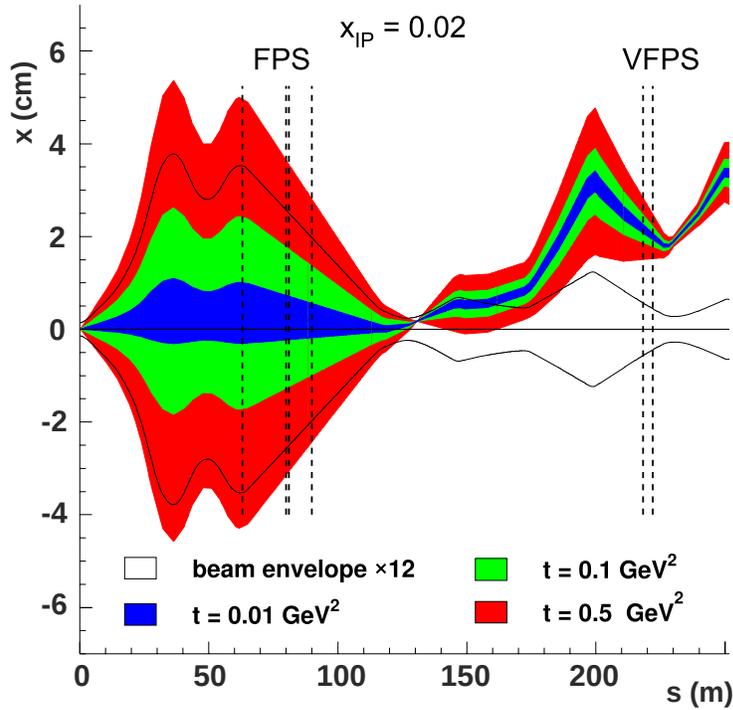} 
\caption{Beam envelope \cite{Astvatsatourov:2014dna} as a function of
  the distance $s$ to the H1 vertex in the $x$ projection, for the $p$
  beam and diffractive protons at $\xpom =0.02$ and $|t|=0.01, 0.1$
  and $0.5$ GeV$^2$.
  The locations of FPS/VFPS stations are indicated by the vertical lines.}
\label{fig_beam_env}
\end{figure}

The VFPS complements the Forward Proton Spectrometer (FPS)
\cite{VanEsch:1999pi}.
The FPS has a wider acceptance in scattered proton energy ($x_{\pom} <
0.1$) but has only limited geometrical acceptance in the azimuthal
angle of the scattered proton (figure~\ref{fig_beam_env}).
In particular at small $|t|<0.2\,\text{GeV}^2$, the VFPS acceptance is
much better than for the FPS.
More than $70\%$ of the diffractive events have $|t|$ smaller than
$0.2\,\text{GeV}^2$.

The Roman pots, which are moved close to the beam as soon as the beam
conditions are sufficiently stable, are equipped with detectors made
of several layers of scintillating fibers with photomultiplier
readout.
The sensitive detector areas are covered by scintillator tiles, the
signals of which are used as a trigger.
The VFPS has high track efficiency ($\sim 96\%$) and low background contamination ($\sim 1\%$).

\subsection{Kinematic Reconstruction}
The observable $x_{\pom}$ is reconstructed by the VFPS from the
relative distance and angle between the track reconstructed between
the two stations and the beam and can be expressed as
\begin{equation}
x_{\pom} = 1 - \frac{E'_p}{E_p}\, ,
\end{equation}
where $E'_p$ is the energy of the leading proton in the VFPS and $E_p$
is the proton beam energy.

The quality of the reconstruction of $x_{\pom}^{\mathrm{VFPS}}$ was
checked using an event sample of elastically produced $\rho$ mesons,
$ep \to e \rho p$.
The $x_{\pom}^{\rho}$ variable reconstructed from the $\rho$ decay
tracks detected in the CTD is compared to $x_{\pom}^{\mathrm{VFPS}}$
determined by the VFPS stations.
The resulting $x_{\pom}^{\mathrm{VFPS}}-\nobreak x_{\pom}^{\rho}$
distributions are found to be in good agreement with Monte Carlo
simulations \cite{Hreus:2008zz}.
The resolution of $x_\pom^{\mathrm{VFPS}}$ was determined to be equal
to $0.0022$ \cite{Astvatsatourov:2014dna}.

The invariant mass $M_X$ of the system $X$ is calculated from all
hadronic objects in the main H1 detector:
\begin{equation}
M_X^2 = \left(\sum_{i \in X} E_i \right)^2 - \left(\sum_{i \in X} \vec{P}_i\right)^2.
\end{equation}
The hadronic final state (HFS) is reconstructed using an energy flow algorithm which combines information from the trackers and calorimeters by avoiding double-counting of energies \cite{Peez:2003zd,Hellwig:2004yp}.

Jets are reconstructed from the hadronic final state objects using the
longitudinally invariant $k_{T}$-jet algorithm \cite{Catani:1992zp}
with a jet distance parameter $R=1.0$ as implemented in the FastJet package
\cite{Cacciari:2011ma}.
The massless $p_T$-recombination scheme is used.
The jet finding algorithm is applied in the $\gamma^{*} p$ frame.
In photoproduction 
this frame is identical to the laboratory frame up
to a Lorentz boost along the beam axis.

The jet properties are studied in terms of the transverse energy of
the leading jet $E_T^{*\mathrm{jet1}}$ in the $\gamma^{*}p$ frame,
of the invariant mass of the
dijet system $M_{12}$ and of the pseudorapidity variables
$\vert\Delta\eta^{\mathrm{jets}}\vert$ 
and $\langle \eta^{\mathrm{jets}}
\rangle$ defined in the laboratory frame, where
\begin{eqnarray}
M_{12}^2 & = & \left(J^{(1)} + J^{(2)}\right)^2, \\
\vert\Delta \eta^{\mathrm{jets}} \vert & = & \left\vert
  \eta^{\mathrm{jet1}} - \eta^{\mathrm{jet2}} \right\vert, \\
\langle \eta^{\mathrm{jets}} \rangle & = & \frac{1}{2}
\left(\eta^{\mathrm{jet1}} + \eta^{\mathrm{jet2}} \right).
\end{eqnarray}
In these definitions, $J^{(1)}$ and $J^{(2)}$ denote the four-momenta of the two leading
jets.

\subsubsection{DIS}
For DIS events the polar angle $\theta_e'$ and energy $E_e'$ of the
scattered positron are measured in the SpaCal calorimeter.
The kinematic reconstruction method introduced in \cite{Adloff:1997sc}
is used
\begin{equation}
y= y_{DA} +  y_e^2 - y_{DA}^2, \qquad Q^2 = \frac{4E_e^2 (1-y)} {\tan^2{\frac{\theta_e'}{2}}}.
\label{yq2_weighted}
\end{equation}
This method interpolates between $y_e$ determined from the scattered
positron alone at larger inelasticity and $y_{DA}$ determined using
the double angle method at low $y$.

The variable $z_{\pom}^{obs}$ is calculated as
\begin{equation}
z_{\pom}^{obs} = \frac{Q^2+M_{12}^2}{Q^2+M_X^2}.
\end{equation}

\subsubsection{Photoproduction}
In the 
$\gamma p$ regime the scattered positron leaves the
interaction undetected. Therefore the inelasticity $y$ is reconstructed
from the hadronic final state
\begin{equation}
y=\frac{\sum_{i \in X} (E_i - P_{z,i})}{2E_e},
\label{y_had}
\end{equation}
where $E_e$ is the initial positron beam energy.

The observables $x_{\gamma}^{obs}$ and $z_{\pom}^{obs}$ are calculated from the
hadronic final state $X$ as
\begin{equation}
x_{\gamma}^{obs} = \frac{\sum_{i \in \mathrm{jets}} (E_i - P_{z,i})}{\sum_{i \in X} (E_i - P_{z,i})} \quad \text{and}\quad   z_{\pom}^{obs} = \frac{\sum_{i \in \mathrm{jets}} (E_i + P_{z,i})}{\sum_{i \in X} (E_i + P_{z,i})},
\end{equation}
where the sums in the numerators run over the leading and the
sub-leading jet, whereas the sums in the denominators include all
objects of the reconstructed hadronic final state.

\subsection{Event Selection}
\label{sec_event_selection}
The analysis is based on a data sample corresponding to an integrated
luminosity of $30\,\text{pb}^{-1}$ for photoproduction and
$50\,\text{pb}^{-1}$ for DIS collected with the H1 detector in the years $2006$ and
$2007$ with proton and positron beam energies of $920\,\text{GeV}$ and
$27.6\,\text{GeV}$, respectively.
The events are triggered on the basis of a coincidence of VFPS signals
from both stations, together with conditions on the charged track
transverse momenta and track topology in the H1 main detector
\cite{Delvax:2010zz}.
%
The trigger efficiency, calculated  using events collected with an
independent trigger condition, was found to be about $80\%$ with
negligible dependence on kinematic quantities.
This efficiency is well reproduced by the H1 trigger simulation after
correcting for an overall normalisation difference of $5\%$.
For the DIS analysis the integrated luminosity 
is increased using the fact that
for most of the DIS events also another trigger based on signals in
the SpaCal has fired.
Only events with a VFPS track in a fiducial volume of high efficiency
are selected \cite{Astvatsatourov:2014dna}.
The reconstructed $z$-coordinate of the event vertex is required to be
within $30\,\text{cm}$ of the mean $z$-position of the interaction point.

The random overlap of $ep$ events with beam-halo protons detected
in the VFPS can constitute a possible background to the VFPS
diffractive data sample.
In such background events the detected proton typically has a small
energy loss, not compatible with the energy loss expected from the
energy deposited in the main H1 detector.
The relative energy loss of the proton detected in VFPS,
$x_\pom^{\mathrm{VFPS}}$, is thus required to be at least $60\%$ of
$x_\pom^{\mathrm{H1}}$ measured in the H1 detector\footnote{The
  variable 
  $x_\pom^{\mathrm{H1}}$ is calculated as
  $x_\pom^{\mathrm{H1}}=\frac{Q^2+M_X^2}{ys}$.},
$x_{\pom}^{\mathrm{VFPS}}/x_{\pom}^{\mathrm{H1}} > 0.6$.
In addition, $x_\pom^{\mathrm{H1}}$ is required to be smaller than $0.04$.
The remaining background contamination after applying the above cuts
is estimated from data by overlaying events without VFPS activity with
VFPS signals recorded independently of any detector activity and is
found to be less than $1\%$ \cite{Astvatsatourov:2014dna}.


The scattered positron candidate of an event is identified as the electromagnetic cluster with the highest transverse momentum being well isolated and having a minimum energy of $8$~GeV.
If such a candidate is absent the event is defined as photoproduction.

For the selection of DIS events in this analysis the positron candidate is required to be detected in the SpaCal.
The energy $E_e'$ and polar angle $\theta_e'$ of the scattered
positron are determined from the SpaCal cluster and the interaction
vertex reconstructed in the CTD.
In order to improve the background rejection, additional requirements
on the transverse cluster radius and lower limit to the positron
energy are imposed \cite{Glazov:2010zza}.
The quantity $\sum_i (E_i - P_{z,i})$ summed over all HFS
particles and the scattered positron, is required to be in the
range $35$-$75$~GeV. For fully reconstructed neutral current
DIS events this quantity is
expected to be twice the positron beam energy ($55.2\,\text{GeV}$) but
is expected to be lower for photoproduction background  where the
scattered positron escapes undetected.
Radiative events where a photon is emitted along the
direction of the incident positron beam, also have a reduced
 $\sum_i (E_i - P_{z,i})$.
 
The leading and the sub-leading jets  are required to have transverse
energies $E_T^{*\mathrm{jet1}}>5.5\,\text{GeV}$ and
$E_T^{*\mathrm{jet2}}>\nobreak 4.0\,\text{GeV}$, respectively.
These cuts are asymmetric in the transverse energy to restrict the
phase space to a region where NLO QCD calculations are reliable
\cite{Klasen:1996b,Frixione:1997a}.
An event is rejected if one of these two jets is outside
of $-1<\eta^{\mathrm{jet1,2}}<2.5$.
Events with $z_\pom^{obs}$ above 0.8 are excluded to improve the reliability
of the comparison between data and theoretical predictions, since the
DPDF are determined with a similar $z_\pom$ restriction.

The DIS events are selected with photon virtualities
$4 \,\text{GeV}^2 < Q^2 < 80  \,\text{GeV}^2$.
Both data samples are restricted to a common $y$ range $0.2<y<0.7$.
In table~\ref{tab_cuts} the definitions of both analysis phase spaces
are summarised. 
The photoproduction and DIS data samples contain $3768$ and $550$
events, respectively.
In addition to the event selection summarised in table~\ref{tab_cuts} an event selection is performed extending the phase space  
in all kinematic variables and other selection requirements to obtain events for an adequate description of migrations at the phase space boundaries.
\begin{table}
\begin{center}
\renewcommand{\arraystretch}{1.3}
\begin{tabular}{|l|ccc|}
\hline
&Photoproduction&$\quad$&DIS\\\hline
 &
$Q^2<2\, \text{GeV}^2$& &  $4\,\text{GeV}^2<Q^2<80\,\text{GeV}^2$\\
\raisebox{1.5ex}[-1.5ex]{Event kinematics}&
\multicolumn{3}{c|}{$0.2<y<0.7$} \\
\hline
&\multicolumn{3}{c|}{$0.010< x_{\pom} < 0.024$} \\ 
Diffractive phase space&\multicolumn{3}{c|}{$|t| < 0.6\,\text{GeV}^2$} \\ 
&\multicolumn{3}{c|}{$z_{\pom} < 0.8$} \\
\hline
&\multicolumn{3}{c|}{$E_T^{*\mathrm{jet1}} > 5.5\, \text{GeV}$} \\
Jet phase space&\multicolumn{3}{c|}{$E_T^{*\mathrm{jet2}} > 4.0\,\text{GeV}$}  \\
&\multicolumn{3}{c|}{$-1<\eta^{\mathrm{jet1,2}}<2.5$} \\
\hline
\end{tabular}
\end{center}
\caption{Phase space of the diffractive dijet VFPS measurement for photoproduction and deep-inelastic scattering.}
\label{tab_cuts}
\end{table}

\subsection{Monte Carlo Simulations}
\subsubsection{Correction to the Data}
%
The Monte Carlo (MC) simulation method is used to correct the data for
effects of detector acceptance, resolution and detector
inefficiencies.
All MC samples are passed through a detailed H1 detector simulation
based on the GEANT program \cite{Brun:1987ma} and are subjected to the
same analysis chain as is used for the data.

Diffractive dijet photoproduction and DDIS events were generated using
the RAPGAP MC generator \cite{Jung:1993gf}.
This generator is based on leading order (LO) parton level QCD matrix
elements with a minimum transverse momentum of the outgoing partons of
$\hat{p}_T^{\mathrm{min}} = 1.7\,\text{GeV}$.
Higher orders are mimicked by initial and final state leading
logarithm parton showers.
Fragmentation is accounted for using Lund string model \cite{Andersson:1983ia}
as implemented in Pythia MC generator \cite{Sjostrand:1995iq}.
The H12006 Fit-B DPDF set \cite{Aktas:2006hy} is used in RAPGAP to
describe the density of partons in the diffractively scattered proton.
In photoproduction a resolved photon contribution is
simulated using the GRV-LO photon distribution function
\cite{Gluck:1991ee}.
In addition to a pomeron exchange contribution also a sub-leading
reggeon contribution is included, corresponding to about $\sim\! 2 \%$
of the total cross section. 
In order to describe the data sufficiently well reweighting
functions are applied in $z_{\pom}^{obs}$, $x_{\pom}$ and $t$.
The reweighting is different for $\gamma p$ and DIS.

%
%
%

\subsubsection{Correction to Theoretical Models}
For comparison of QCD calculations with the diffractive measurements,
it is necessary to convert the calculated NLO parton level cross
sections to the level of stable hadrons by evaluating effects due to
hadronisation, fragmentation
and the influence of pomeron or photon remnants.
The RAPGAP MC generator is used to compute the required hadronisation
correction factors for the diffractive dijet calculations.
These factors are defined for each measured data point by

\begin{equation}
1+ \delta_{\mbox{\tiny hadr}}^i = \frac{\sigma_{i}^{\mathrm{hadr}}}{\sigma_{i}^{\mathrm{part}}},
\end{equation}
where the $\sigma_{i}^{\mathrm{hadr}}$ ($\sigma_{i}^{\mathrm{part}}$)
are the bin-integrated MC cross sections at hadron level (parton
level) in a given bin $i$.
They reduce the predicted NLO parton level cross sections by typically
$\sim\! 9\%$ in photoproduction and enhance the cross sections by
typically $\sim\! 2\%$ in DIS.
In photoproduction the hadronisation correction factor is particularly
large at the second highest $x_\gamma^{obs}$ bin, where contributions with
$x_\gamma^{obs} \sim 1$ at parton level migrate to lower values due to
hadronisation effects.
The hadronisation corrections have uncertainties of $3\%$ \cite{Aaron:2010su}.
The hadronisation corrections determined here are applicable to NLO QCD
predictions, since a good agreement in shape of the parton level
predictions of the MC to the NLO calculations is observed.

In the DIS analysis, the RAPGAP MC generator is also used to correct
the measured data for QED radiation effects. The radiative corrections are defined as
\begin{equation}
1+ \delta_{\mbox{\tiny rad}}^i = \frac{\sigma_{i}^{\mathrm{nrad}}}{\sigma_{i}^{\mathrm{rad}}},
\end{equation}
  where
$\sigma_{i}^{\mathrm{rad}}$ ($\sigma_{i}^{\mathrm{nrad}}$) denote the bin
integrated cross sections obtained from RAPGAP when run with (without)
simulating QED radiation.
The term $\delta_{\mbox{\tiny rad}}^i$ is on average compatible with
zero with a standard deviation of $4\%$ within the phase space.
Radiative corrections in photoproduction are found to be negligible.

For the comparison with the measurement, the NLO
QCD predictions are scaled down by a factor of 0.83
\cite{Aaron:2010aa} 
to account for the contributions from proton dissociation ($M_Y<1.6$~GeV) absent in the current analysis but
included in the extraction of the  H12006 Fit-B DPDF set from the inclusive data \cite{Aktas:2006hy}.

\subsection{Cross Section Measurement}
\label{sec_creoss_section}
In order to correct for detector effects, the dijet cross sections
are calculated at the level of stable hadrons using a matrix unfolding
method \cite{doi:1686351,tikhonov}.
The detector response is described by a matrix $\mathbf{A}$ determined
from the RAPGAP simulation.
It relates the expected vector of event counts, $\langle
\vec{y}_\mathrm{rec} \rangle$, to the true event count vector,
$\vec{x}_\mathrm{true}$, on the level of stable hadrons via the formula
$\langle \vec{y}_\mathrm{rec}
\rangle=\mathbf{A}\vec{x}_\mathrm{true}$.

In order to control migrations
at the phase space boundaries also the neighbouring parts
of the analysis phase space are taken into account.
Of these, the migrations caused by events in which jets have low
transverse momenta $E_T$, high $x_{\pom}$ or low $y$ are most
important.
Similar unfolding techniques have been applied in other jet-based analyses 
\cite{Aaron:2011mp,boris,daniel}.

An estimator of the true-level event count $\vec{x}_\mathrm{true}$ is
obtained by minimising  a $\chi^2$ function (\ref{eq_unf}) with
respect to $\vec{x}_\mathrm{true}$
\begin{equation}
\chi^2 = \chi_A^2 + \tau^2 \chi_L^2 = \frac{1}{2} (\vec{y}_\mathrm{rec} - \mathbf{A}\vec{x}_\mathrm{true})^T \mathbf{V}^{-1} (\vec{y}_\mathrm{rec} - \mathbf{A}\vec{x}_{\mathrm{true}}) + \tau^2 (\vec{x}_{\mathrm{true}} - \vec{x}_b)^T \mathbf{L}^T \mathbf{L} (\vec{x}_{\mathrm{true}} - \vec{x}_b),
\label{eq_unf}
\end{equation}
where the matrix $\mathbf{V}$ is the covariance of data.
The term $\chi_A^2$ is a measure of the agreement between
$\mathbf{A}\vec{x}_{\mathrm{true}}$ and $\vec{y}_{\mathrm{rec}}$,
where $\vec{y}_{\mathrm{rec}}$ is the vector of events counts after
subtracting residual background contributions.
The regularisation term $\tau^2\chi_L^2$ suppresses large fluctuations
of $\vec{x}_\mathrm{true}$.
The type of the regularisation is defined by the matrix
$\mathbf{L}$.
In this paper, $\mathbf{L}$ is set to the unity matrix.
The vector $\vec{x}_B$ defines a bias for the regularisation term,
taken from the RAPGAP prediction.
The value of the regularisation parameter $\tau$ is chosen using the
$L$-curve method as described elsewhere \cite{doi:1686351}.

The bin-integrated cross section for each data point is given by

\begin{equation}
\sigma_i = \frac{x_\mathrm{true}^i}{\mathcal{L}}\, \left(1+\delta_{\mbox{\tiny rad}}^i\right)
\end{equation}
where $\mathcal{L}$ is the integrated luminosity of the data. The
radiative corrections $\delta_{\mbox{\tiny rad}}^i$ are non-zero only
for the DIS case.

\subsection{Systematic Uncertainties of the Measured Cross Section}
For each source of systematic uncertainty, a separate response matrix
$\mathbf{A}$ is filled and the difference to the nominal matrix
$\mathbf{A}$ is propagated through the unfolding procedure.
All these individual contributions of systematic uncertainties are
then added in quadrature for each bin to obtain the total systematic
uncertainty.
The following systematic effects are studied:
\begin{description}
\item[VFPS calibration]  
  The primary source of the VFPS systematic uncertainties is related to an
  uncertainty of the $x$ and $y$ global track coordinates\footnote{The
    global track coordinates are reconstructed by linking the local
    tracks of the two VFPS stations.} with respect to the beam.
  The actual beam position is measured with help of a beam position
  monitor \cite{Astvatsatourov:2014dna,Schutte:1987qn} which has a
  precision of $160\,\mu\text{m}$ in $x$ and $120\,\mu \text{m}$ in $y$.
  The horizontal coordinate $x$ has an additional uncertainty originating
  from the VFPS calibration procedure, tied to the reconstruction of
  $x_\pom$ in the main H1 detector.
  The resulting $x$-coordinate uncertainty is $250\,\mu\text{m}$.

The time variation of the  beam-tilt in $x$ and $y$ introduces an uncertainty of 
$8\,\mu \text{rad}$ for the $x$-tilt
  and $6\,\mu \text{rad}$ for the $y$-tilt.

  More details on the VFPS reconstruction and its precision are given in
  \cite{Astvatsatourov:2014dna}.
  In total, all sources of the VFPS uncertainties affect the integrated
  cross section by $5.5\%$ in $\gamma p$ 
  and typically $3.7\%$ in DIS.

\item[Positron reconstruction] In the DIS analysis the uncertainties
  of the measured positron energy $E'_e$ ($1\%$) and angle $\theta'_e$
  ($1\,\text{mrad}$) in the SpaCal calorimeter lead to an uncertainty
  of the total cross section of $0.4\%$ and $0.7 \%$, respectively.

\item[Energy scale] The uncertainty of the hadronic final state energy
  calibration is $2\%$ \cite{Salek:2010ifa}.
  It affects the total cross section by $\pm 7.6\%$ for
  photoproduction and by $\pm 6.1\%$ for DIS.

\item[Model uncertainties] The influence of the MC model used to
  unfold the cross sections is studied by varying the kinematic
  distributions of the RAPGAP MC generator within certain limits while
  maintaining an acceptable description of the data.
  For this purpose the shape of the kinematic distributions in $E_T^{*\mathrm{jet1}}$, $x_\pom$, $z_\pom$,
  $x_\gamma$, $y$, $t$ and $Q^2$ are altered by applying multiplicative weights of 
$(E_T^{*\mathrm{jet1}})^{\pm 0.4}$,
$x_{\pom}^{\pm 0.2}$,
$z_{\pom}^{\pm 0.3}$,
$x_{\gamma}^{\pm 0.3}$,
$y^{\pm 0.3}$,
$e^{\pm t}$ and 
$(Q^2+0.1\,\mathrm{GeV}^2)^{\pm 0.2}$, respectively.
The largest resulting uncertainties arise from variation of the shape
in $t$ ($4.5\%$ in $\gamma p$ and $3.3\%$ in DIS)
and $E_T^{*\mathrm{jet1}}$ ($3.5\%$ in $\gamma p$ and $3.0\%$ in DIS).
The integrated cross section uncertainty due to model dependence is
$7\%$ in $\gamma p$ and $5\%$ in DIS.

\item[Normalisation uncertainties]  
  The following sources of systematic normalisation errors are considered:
  \begin{itemize}
  \item The VFPS track reconstruction efficiency is known to within
    $2.5\%$ \cite{Astvatsatourov:2014dna}.
  \item The VFPS background originating from interactions of beam
    particles with the residual gas, producing a proton signal in the VFPS
    in 
    accidental coincidence with a dijet event in the main H1
    detector is less than $1\%$ and is treated as a normalisation
    uncertainty \cite{Astvatsatourov:2014dna}.
  \item The integrated luminosity of the VFPS triggered data is known
    to within $3\%$ \cite{Aaron:2012kn}.
  \item The trigger efficiency has an uncertainty of $5\%$. 
  \end{itemize}
  The resulting total normalisation uncertainty amounts to $6\%$.

\end{description}

\begin{figure}[p]
\includegraph[width=1.0\textwidth]{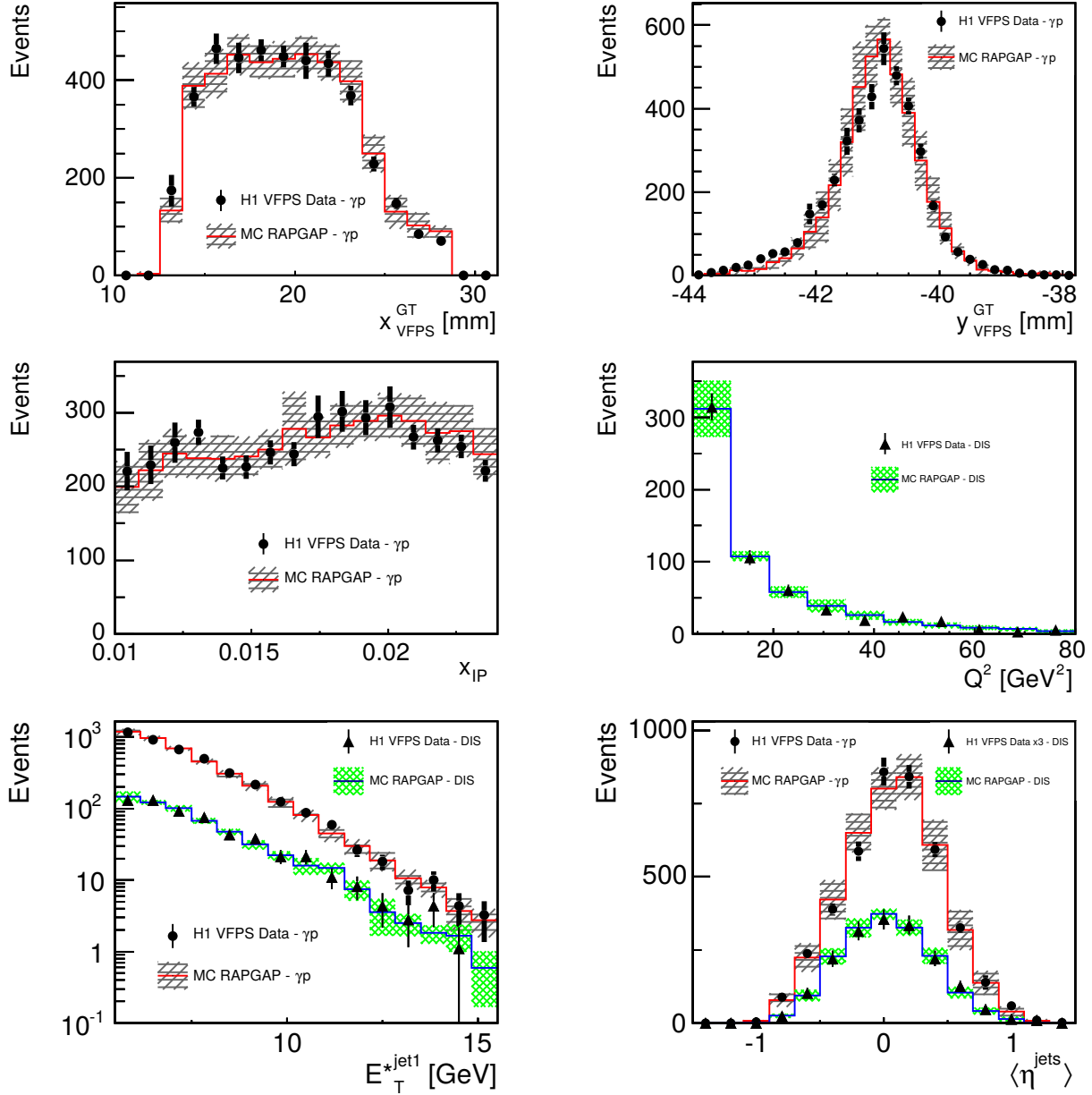} 
\caption{Comparison of the photoproduction data (dots) and DIS data
  (triangles) with the reweighted RAPGAP MC simulation (solid line)
  as a function of coordinates $x$ and $y$ in VFPS, $\xpom$, $Q^2$,
  $E_T^{*\mathrm{jet1}}$ and $\langle\eta^{\mathrm{jets}}\rangle$.
 The systematic uncertainties are shown as bands on the histograms.
 For better visibility the $\langle\eta^{\mathrm{jets}}\rangle$ DIS
 distribution is multiplied by a factor 3.
}
\label{fig_control_plots}
\end{figure}

Figure \ref{fig_control_plots} displays the distributions of the $x$- and $y$-coordinates of global
tracks in the VFPS, $\xpom$ as measured by the VFPS,
$Q^2$ for the DIS selection and the jet variables 
$E_T^{*\mathrm{jet1}}$ and $\langle\eta^{\mathrm{jets}}\rangle$ in
comparison to the MC distributions after reweighting 
and normalising to the data.
In all cases the data are well described in shape within systematic errors.

\section{Results}
\subsection{Integrated Photoproduction and DIS Cross Sections}
\begin{table}[t]
\begin{center}
%
%
%
%
%
\renewcommand{\arraystretch}{1.2}
\setlength{\tabcolsep}{0.1em}
\newcommand{\DATASTATSYSTOT}[6]{$#3$&$\pm#4$ (stat)&$\pm#5$ (syst) & }
\newcommand{\NLO}[4]{#1&#2 {\small(scale)}&#3 {\small(DPDF)}&#4 {\small(hadr)}}
\newcommand{\RAPGAP}[1]{$#1$& & & }
\newcommand{\RATIODATATHEORY}[3]{#1#2 (data)#3 (theory)}
\begin{tabular}{l|rrrr|rrrr}\hline
 & \multicolumn{4}{c|}{PHP} & \multicolumn{4}{c}{DIS} \\ \hline\hline
Data [pb] &
\DATASTATSYSTOT{ }{ }{237}{14}{31}{0.906}
 & 
\DATASTATSYSTOT{ }{ }{30.5}{1.6}{2.8}{1.022}
 \\
NLO QCD [pb] &
\NLO{$430$}{${}^{+172}_{-98}$}{${}^{+48}_{-61}$}{$\pm13$}
 & 
\NLO{$28.3$}{${}^{+11.4}_{-6.4}$}{${}^{+3.0}_{-4.0}$}{$\pm0.8$}
 \\
RAPGAP [pb] &
\RAPGAP{180}
 & 
\RAPGAP{18.0}
 \\ \hline\hline
Data/NLO &
\multicolumn{4}{c|}{\RATIODATATHEORY{$0.551$}{$\pm0.078$}{${}^{+0.230}_{-0.149}$}}
 & 
\multicolumn{4}{c}{\RATIODATATHEORY{$1.08$}{$\pm0.11$}{${}^{+0.45}_{-0.29}$}}
 \\ \hline
\end{tabular}
\end{center}
\caption{Integrated $e^{+}p$ diffractive dijet cross sections in
  $\gamma p$ and DIS compared to NLO QCD calculations
  using the H12006 Fit-B DPDF set.
  The measured cross sections are gresented with statistical and
  systematical uncertainties.
  For the theoretical predictions the uncertainties from scale
  variations, from the H12006 Fit-B DPDFs and from the hadronisation corretions are given.
  The predictions from RAPGAP are also shown.
%
%
  The ratios data/NLO are given in the last row.}
\label{tab_xsectot}
\end{table}
The integrated $e^{+}p$ diffractive dijet cross sections in the $\gamma p$
and in the DIS regime measured in the kinematic range defined in
table~\ref{tab_cuts}  are presented together with NLO QCD and RAPGAP predictions
in table \ref{tab_xsectot}.
The total theoretical uncertainty is calculated by using the sign improved
quadratic sum of DPDF eigenvectors \cite{boris}, scale and
hadronisation uncertainties.
In the DIS regime, the theoretical expectation agrees with the measurement
within uncertainties.
This confirms the observation made in previous measurements
\cite{Aktas:2007bv,Aaron:2011mp,Chekanov:2007aa,boris}.
In contrast, the integrated diffractive dijet cross section in
photoproduction is overestimated by the NLO QCD theory by almost a
factor of two, with considerable theory uncertainty.
This observation agrees with the results of previous H1 analyses in a
similar kinematic range \cite{Aaron:2010su,Aktas:2007hn}, based on
different data sets and using different experimental techniques to
select diffractive events. To conclude, the integrated NLO QCD 
cross section predictions are in
disagreement with three independent H1 measurements of diffractive
dijet photoproduction.
The MC RAPGAP, based on leading order matrix elements and parton
showers, fails to describe the integrated cross sections
both in DIS and in photoproduction.

\subsection{Diffractive Dijet Production in DIS}
\begin{figure}[p]
\begin{center}
\includegraph[width=0.4\textwidth]{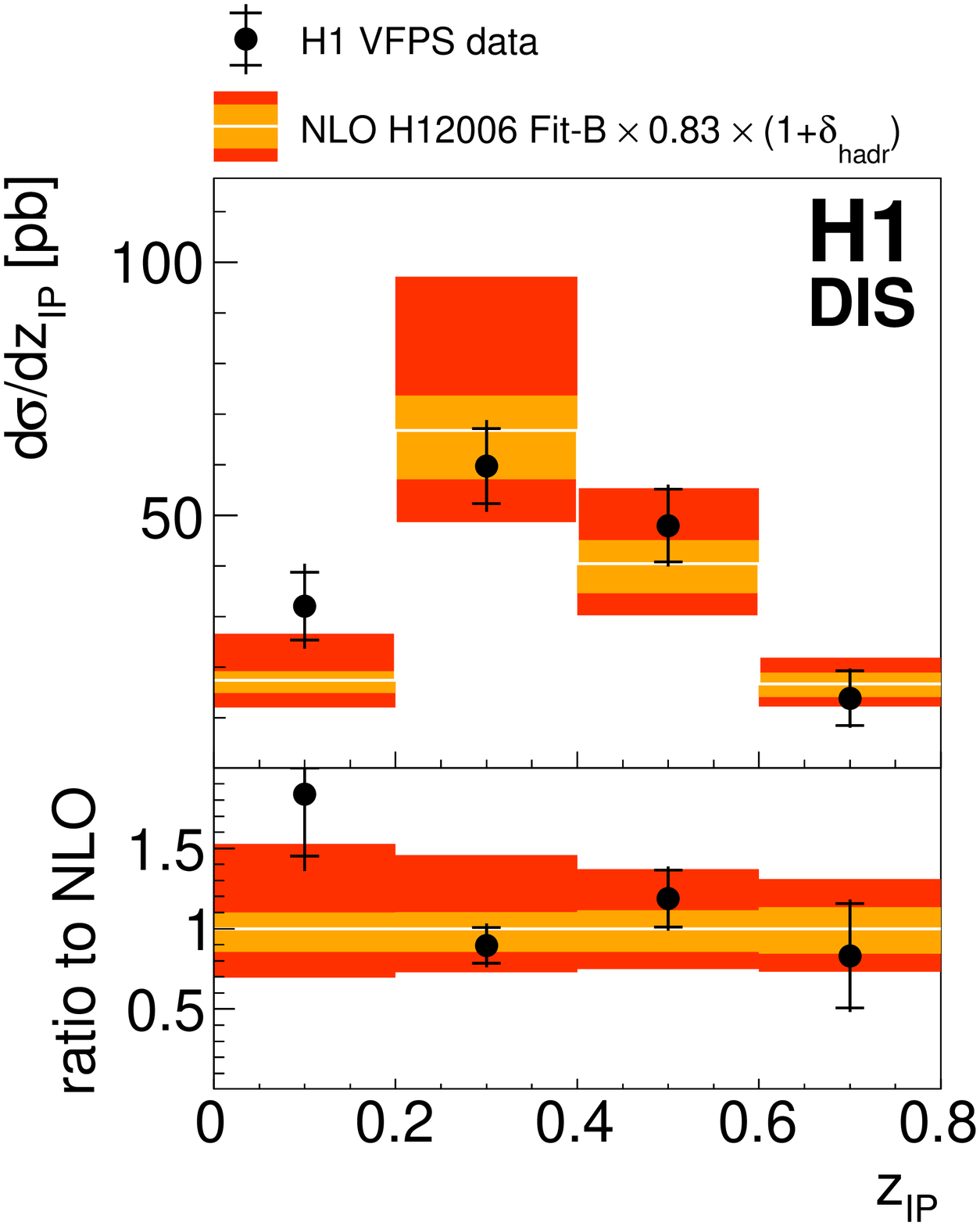}
\includegraph[width=0.4\textwidth]{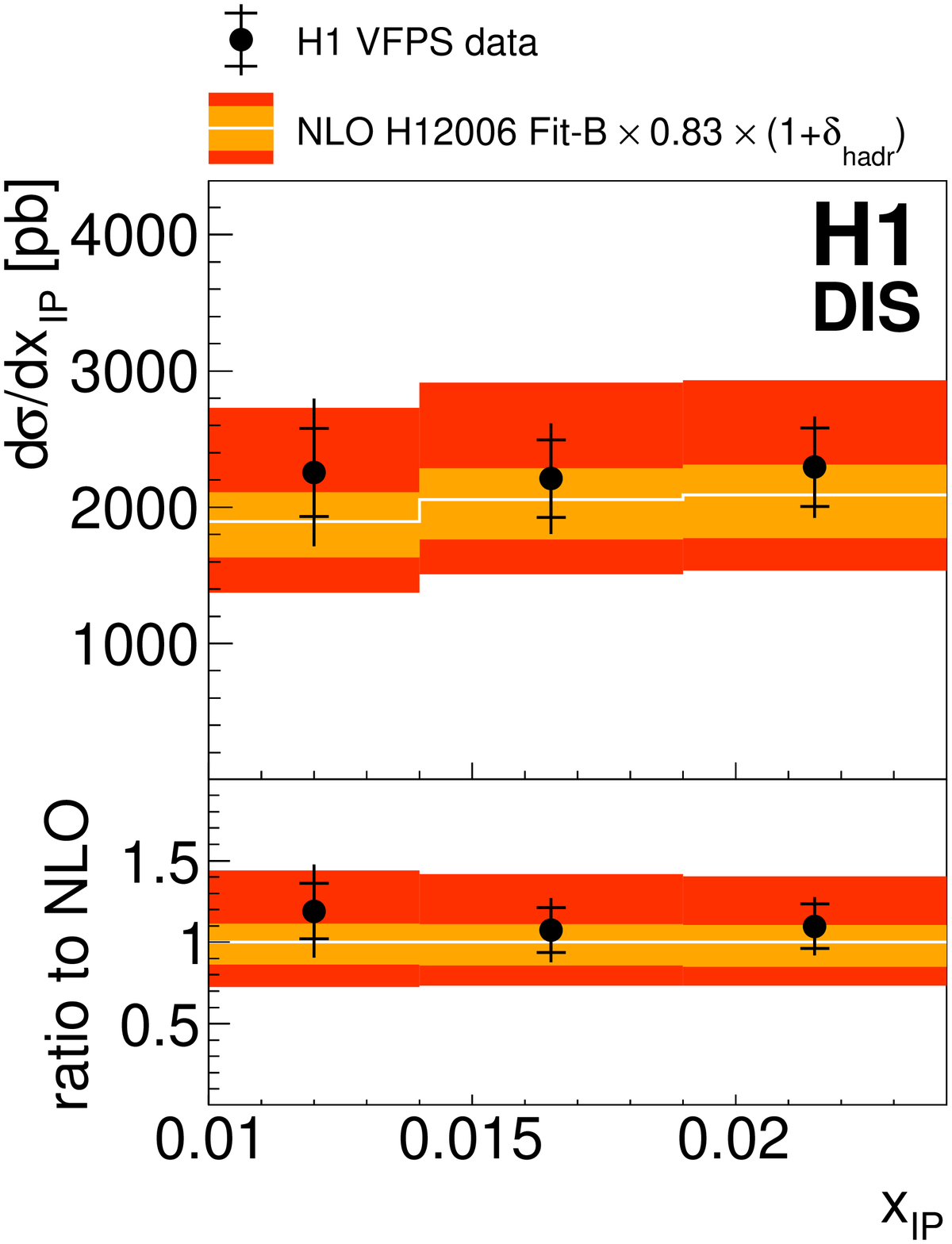}
\includegraph[width=0.4\textwidth]{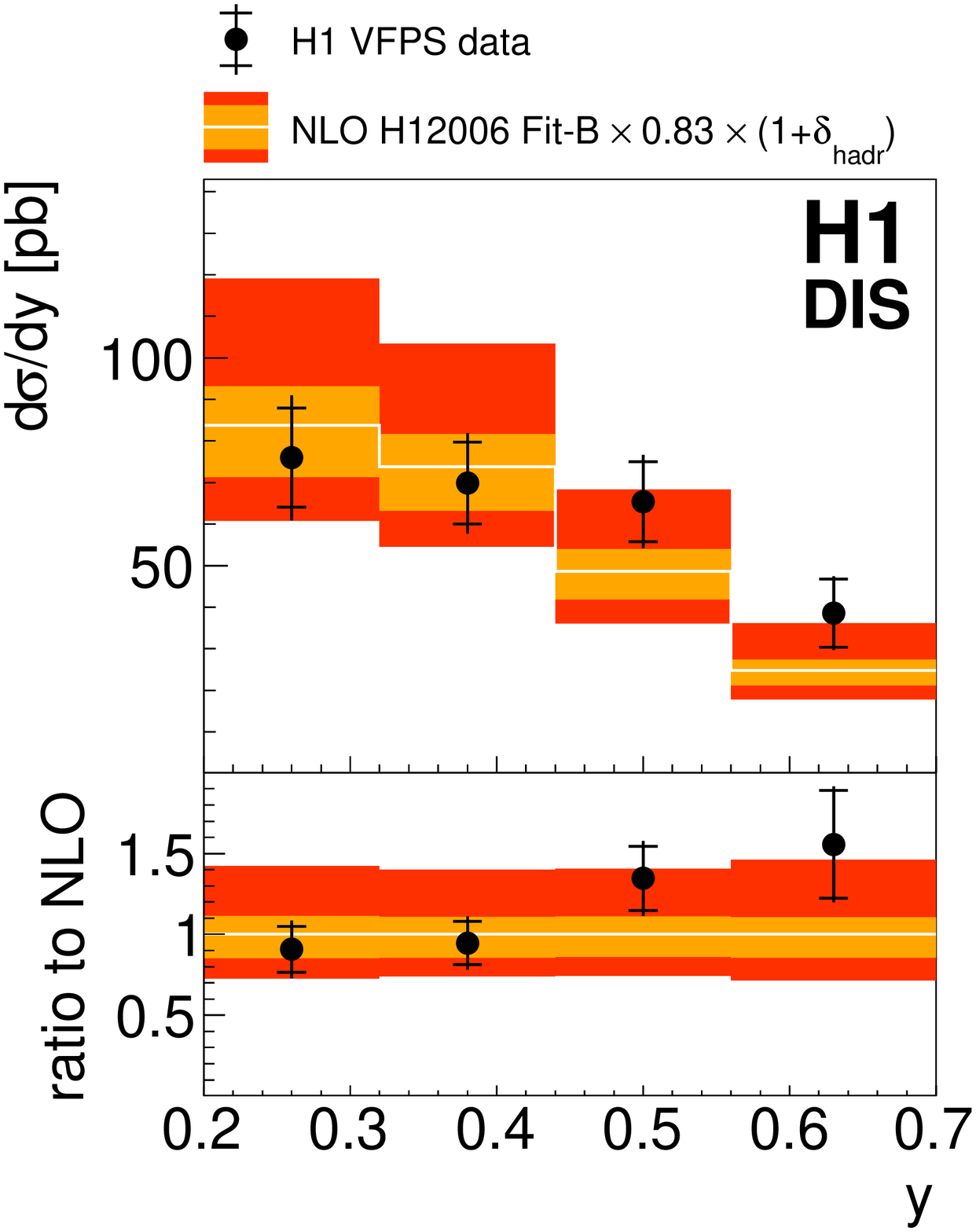}
\includegraph[width=0.4\textwidth]{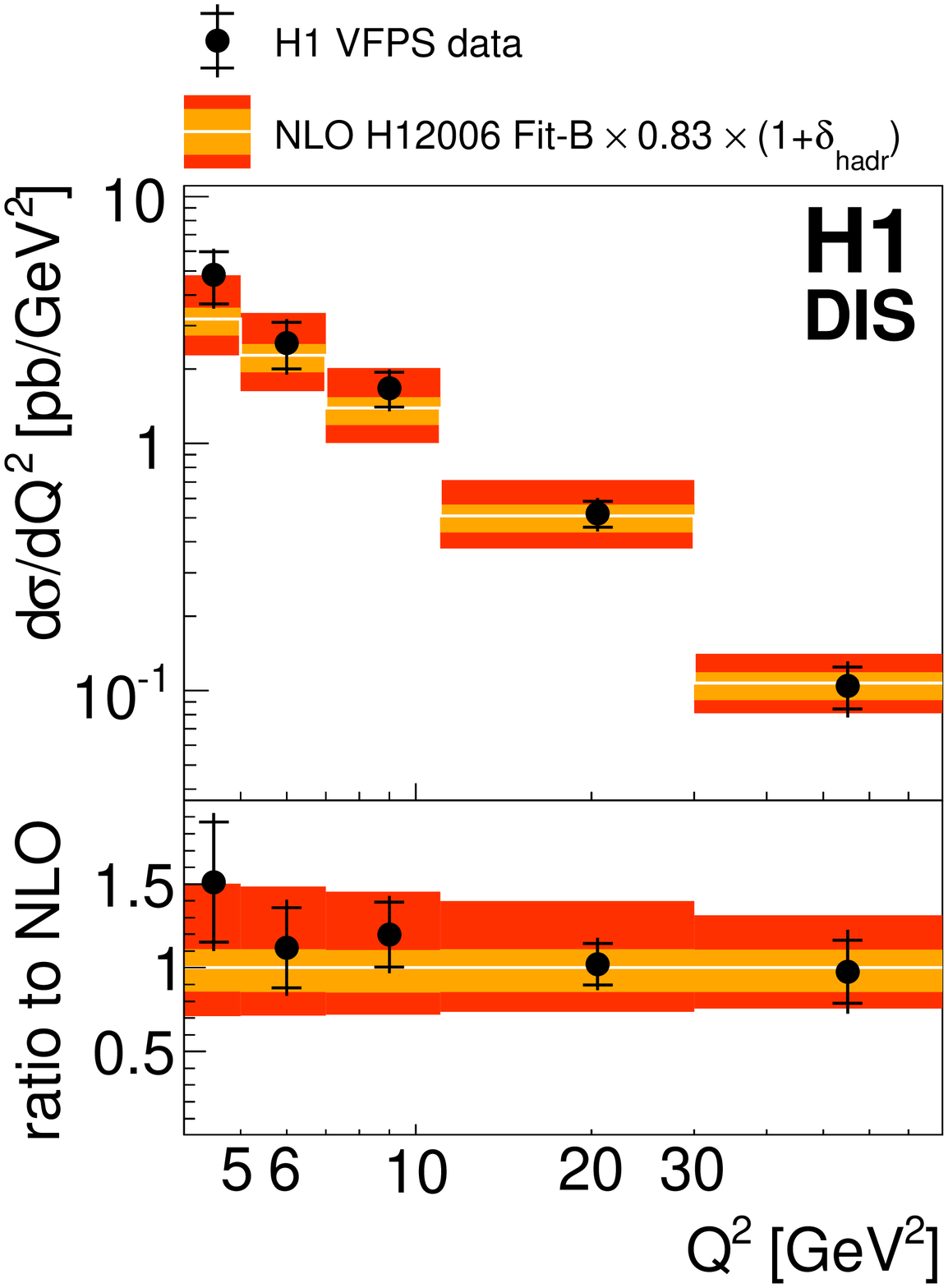}
\end{center}
\caption{Diffractive dijet DIS cross sections differential in
  $z_{\pom}$, $x_{\pom}$, $y$ and $Q^2$.
  The inner error bars represent the statistical errors.
  The outer error bars indicate the statistical and systematic errors
  added in quadrature.
  The overall normalisation uncertainty of $6\%$ is not shown.
  NLO QCD predictions based on the H12006 Fit-B DPDF set,
  corrected to the level of stable
  hadrons, are shown as a white line.
  They are scaled by a factor $0.83$ to account for
  contributions from proton-dissociation which are present in the DPDF
  fit but not in the data.
  The inner, light shaded band indicates the size of the DPDF uncertainties
  and hadronisation corrections added in quadrature.
  The outer, dark shaded band indicates the total NLO uncertainty, also
  including scale variations by a factor of $0.5$ to $2$.
  For each variable, the cross section is shown in the upper panel,
  whereas the ratio to the NLO prediction is shown in the lower panel.
}
\label{fig_DisXsec1}
\end{figure}

\begin{figure}[p]
\begin{center}
\includegraph[width=0.4\textwidth]{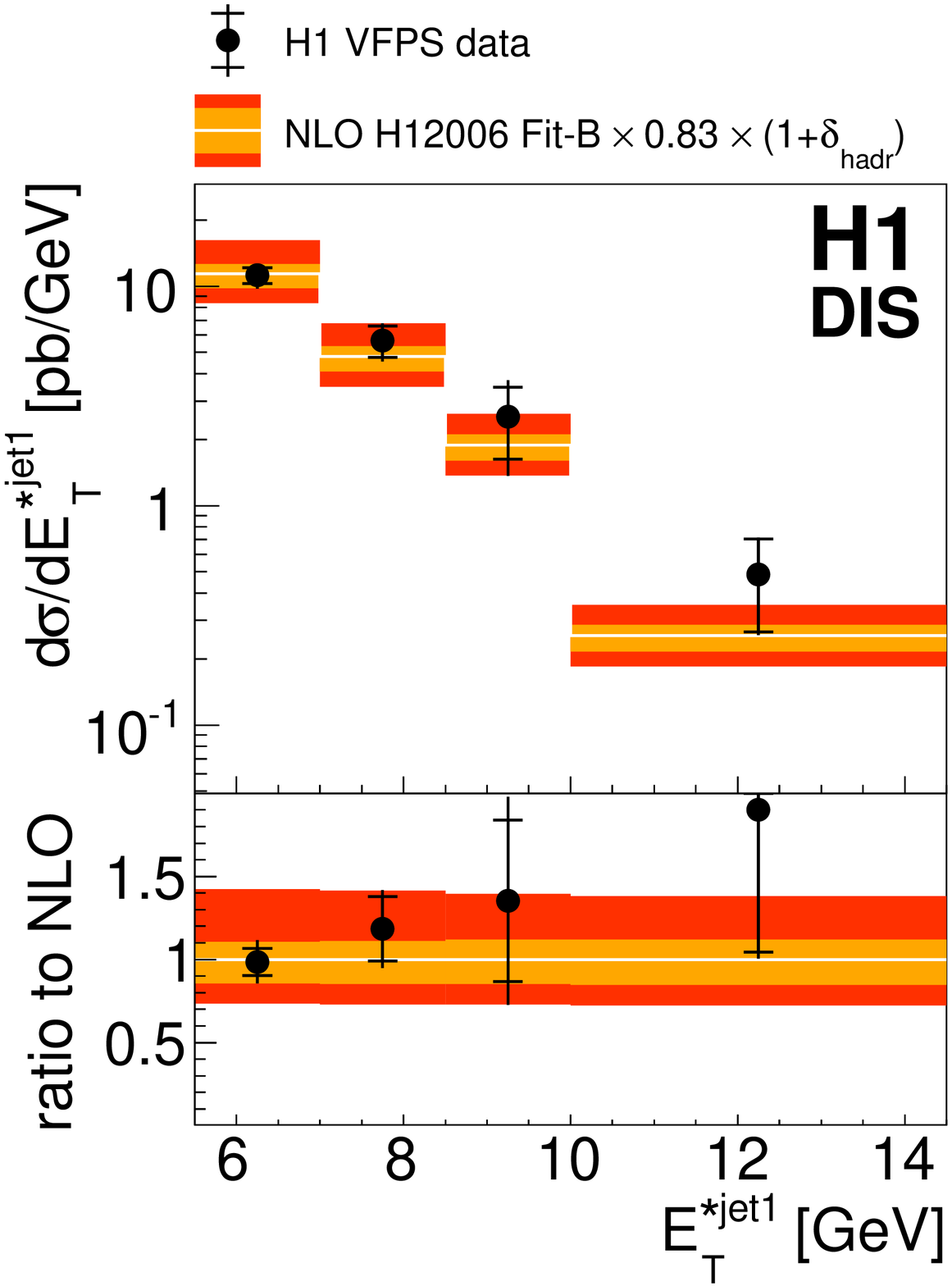}
\includegraph[width=0.4\textwidth]{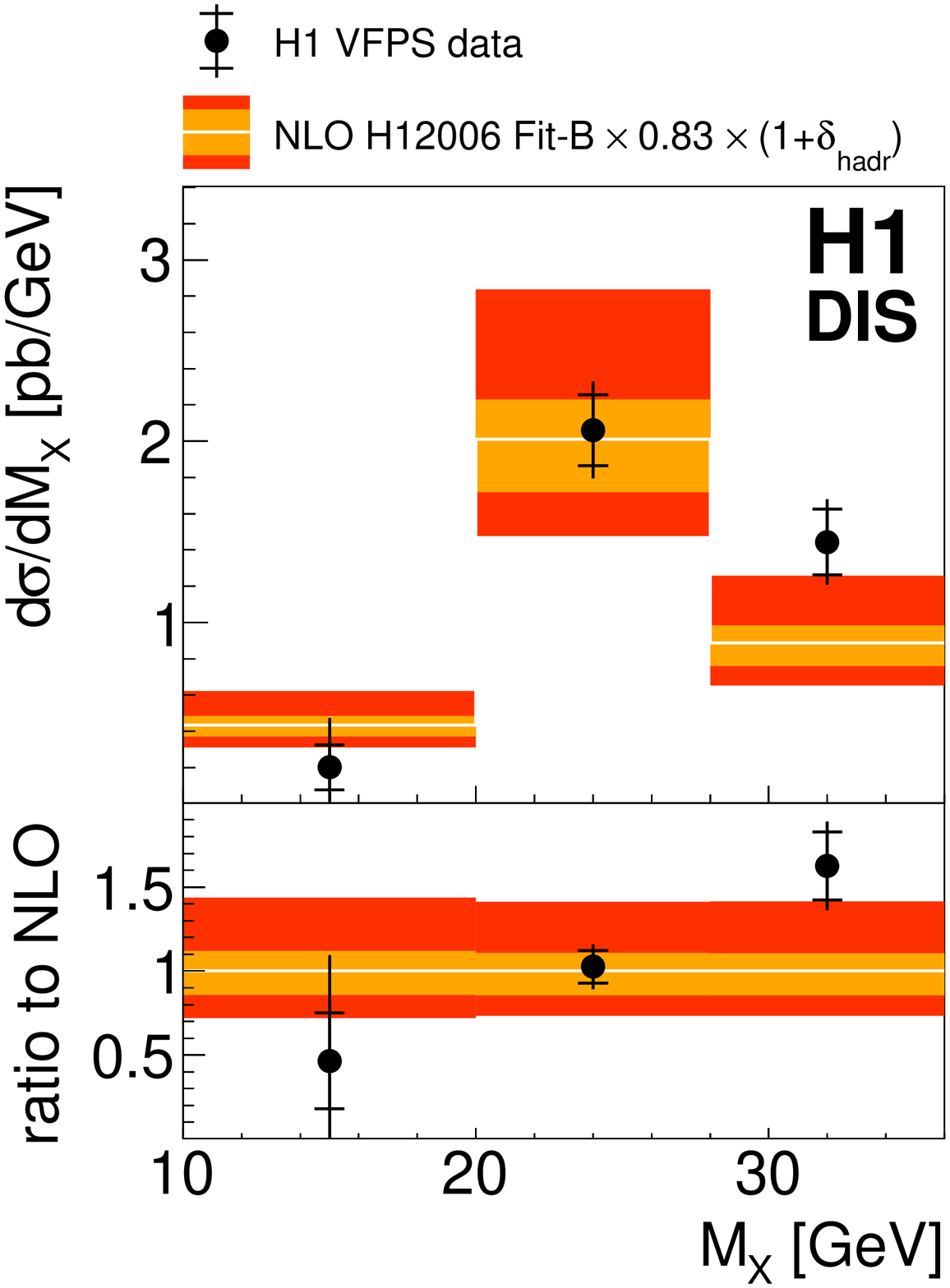}
\includegraph[width=0.4\textwidth]{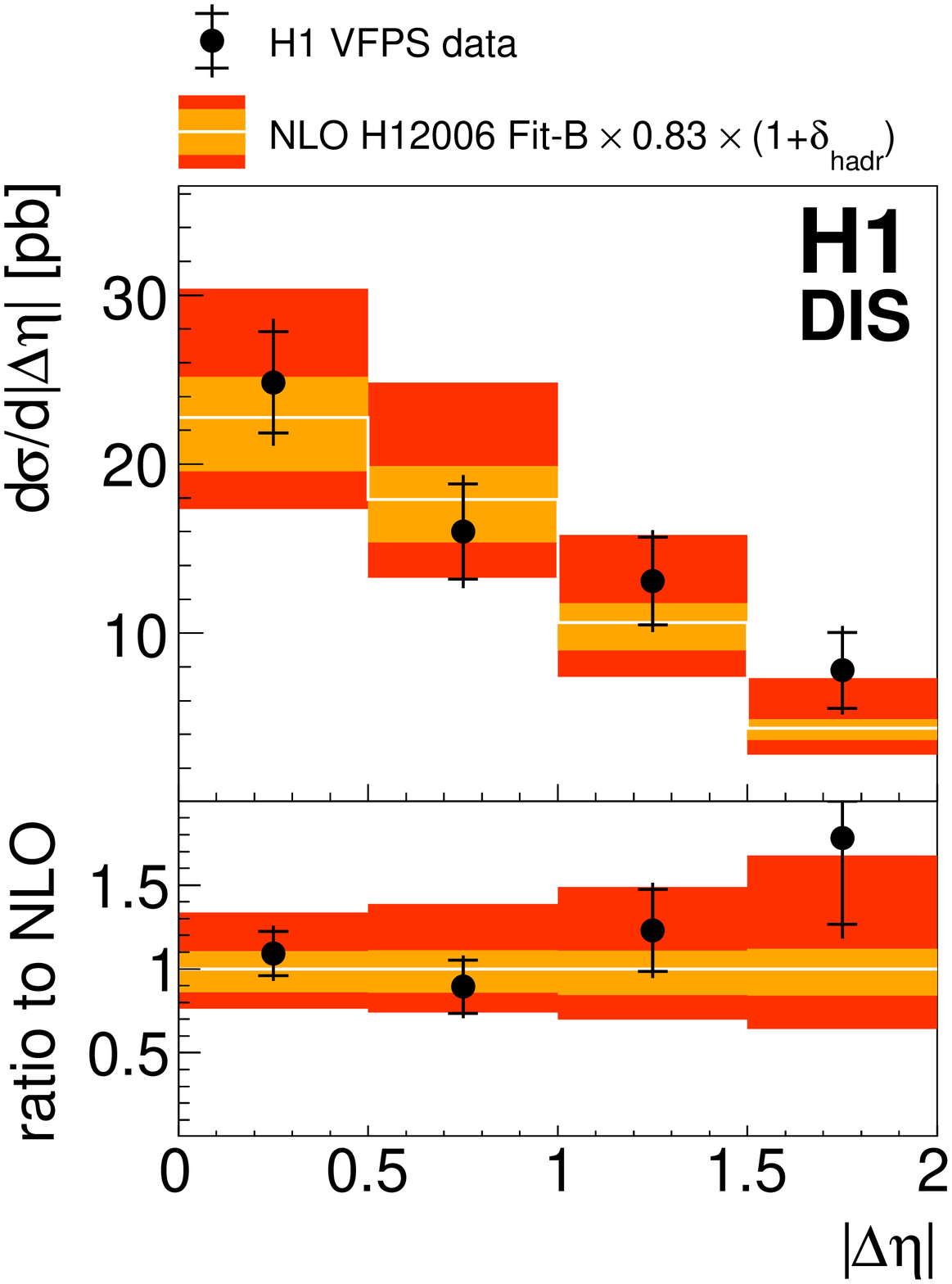}
\includegraph[width=0.4\textwidth]{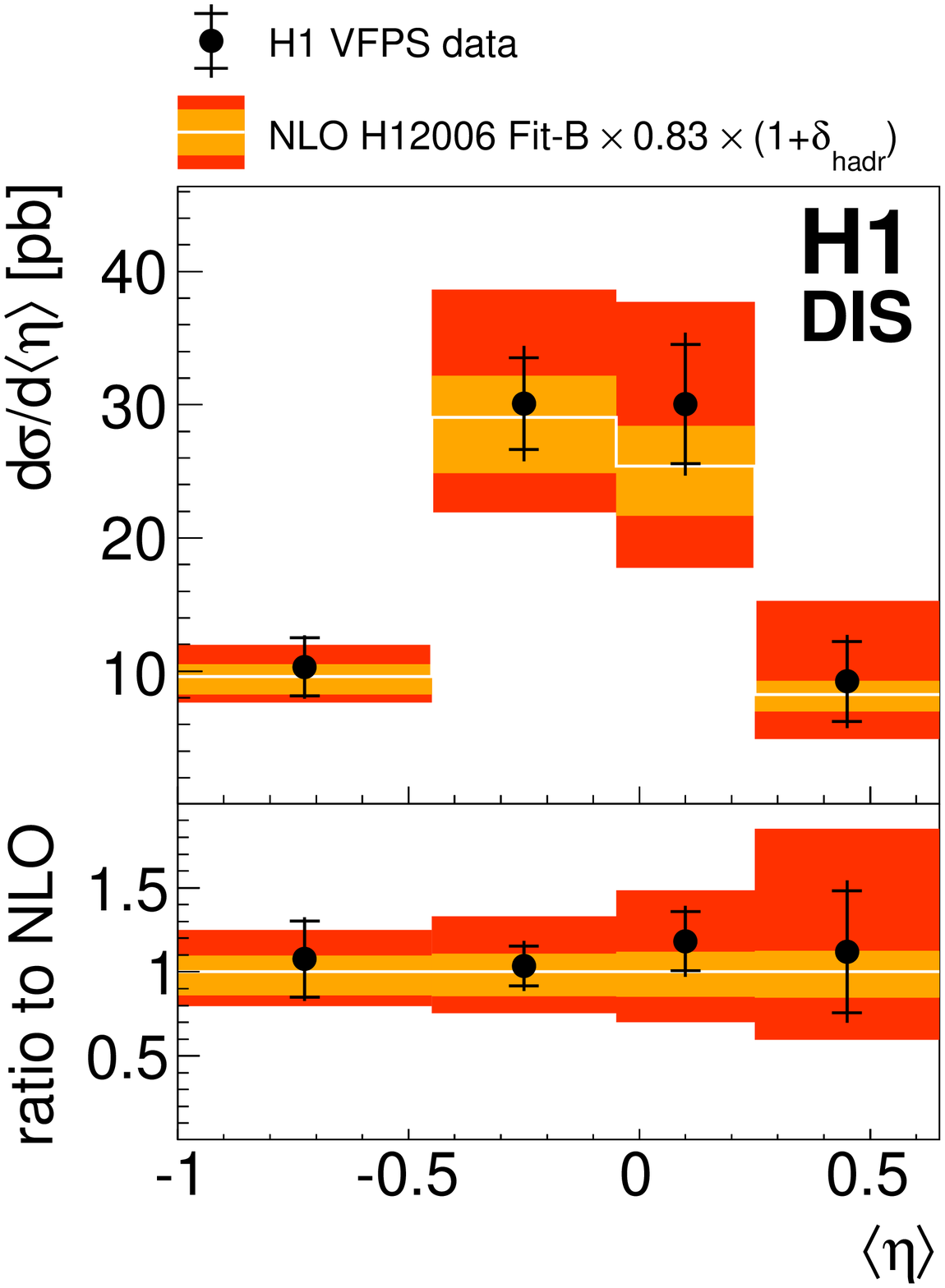}
\end{center}
\caption{Diffractive dijet DIS cross sections differential in
  $E_T^{*\mathrm{jet1}}$,
  $M_X$,
  $|\Delta \eta^{\mathrm{jets}}|$
  and
  $\langle\eta^{\mathrm{jets}}\rangle$.
  The inner error bars represent the statistical errors.
  The outer error bars indicate the statistical and systematic errors
  added in quadrature.
  Further details are given in the caption of figure \ref{fig_DisXsec1}.
}
\label{fig_DisXsec2}
\end{figure}

The measured differential DIS cross sections as a function of
$z_{\pom}^{obs}$, $x_{\pom}$, $y$, $Q^2$ are given in
table~\ref{tab_DisXsec1} and are shown in figure~\ref{fig_DisXsec1}
together with the NLO QCD predictions.
In table \ref{tab_DisXsec2} and figure~\ref{fig_DisXsec2} the
differential cross sections in DIS are shown as a function of
$E_T^{*\mathrm{jet1}}$, $\langle\eta^{\mathrm{jets}}\rangle$, $\vert\Delta
\eta^{\mathrm{jets}} \vert$ and $M_X$.
The NLO QCD predictions are in good agreement with the measurements 
within data and theory uncertainties.

The shapes of the NLO predictions are tested using the ratio of data
to prediction.
A somewhat different shape is observed for data and theory as a
function of $Q^2$, however the deviations are covered by the
uncertainties.
Resolved photon \cite{sedlak} and higher twist
contributions \cite{highertwist} are expected to change the DIS
cross sections at small $Q^2$.
The predicted shape in $y$ also differs from the observation,
such that at high $y$ smaller cross sections are predicted than
observed.
Similar shape deviations in $Q^2$ and $y$ have also been observed
in a recent measurement of diffractive dijet production based on
a large rapidity gap selection \cite{boris}.
The cross section as a function
of $E_T^{*\mathrm{jet1}}$ 
is observed to be slightly harder than predicted by theory, although
still in agreement within uncertainties.

\subsection{Diffractive Dijet Production in Photoproduction}
\label{sec_dpho_results}
\begin{figure}[p]
\begin{center}
\includegraph[width=0.4\textwidth]{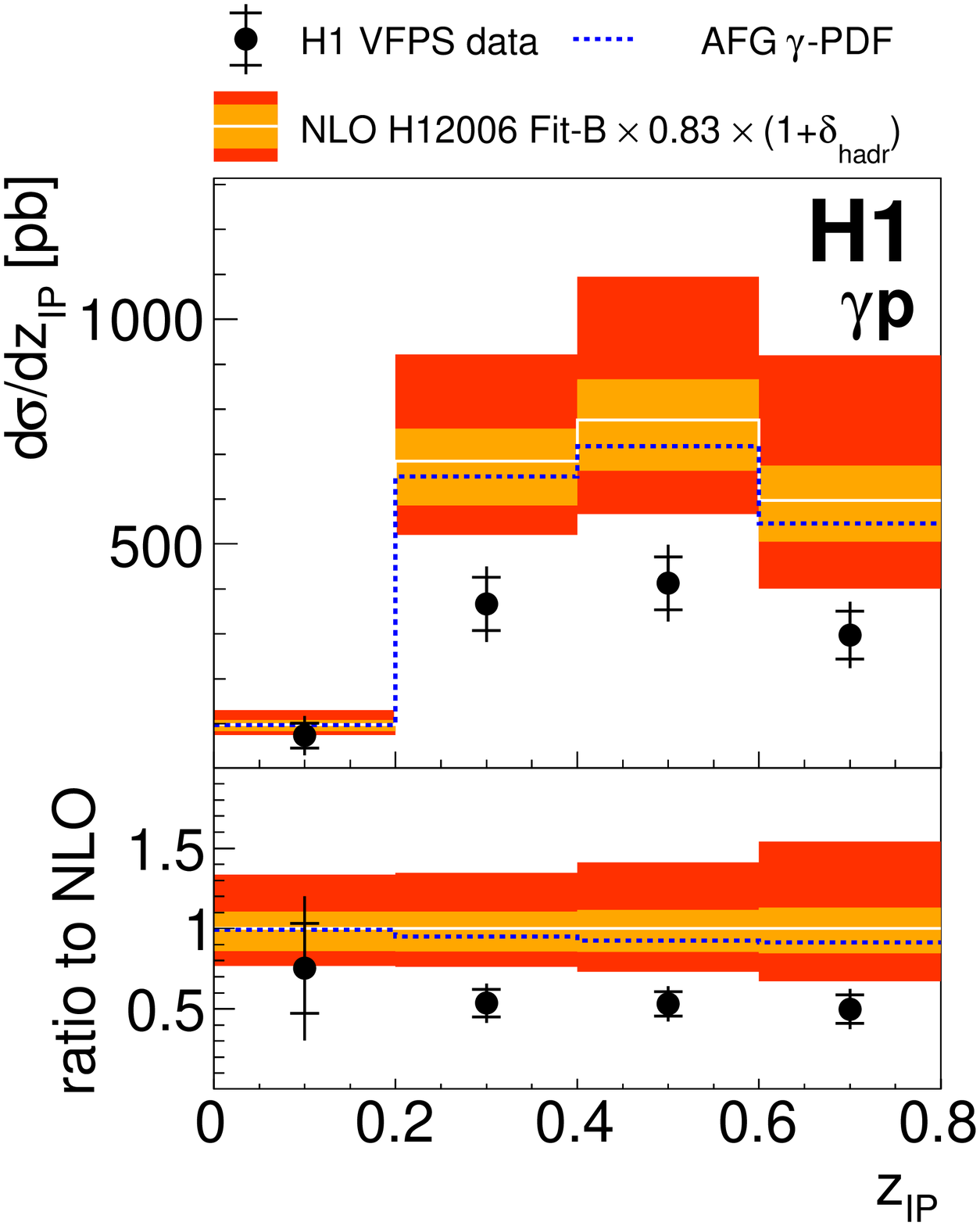}
\includegraph[width=0.4\textwidth]{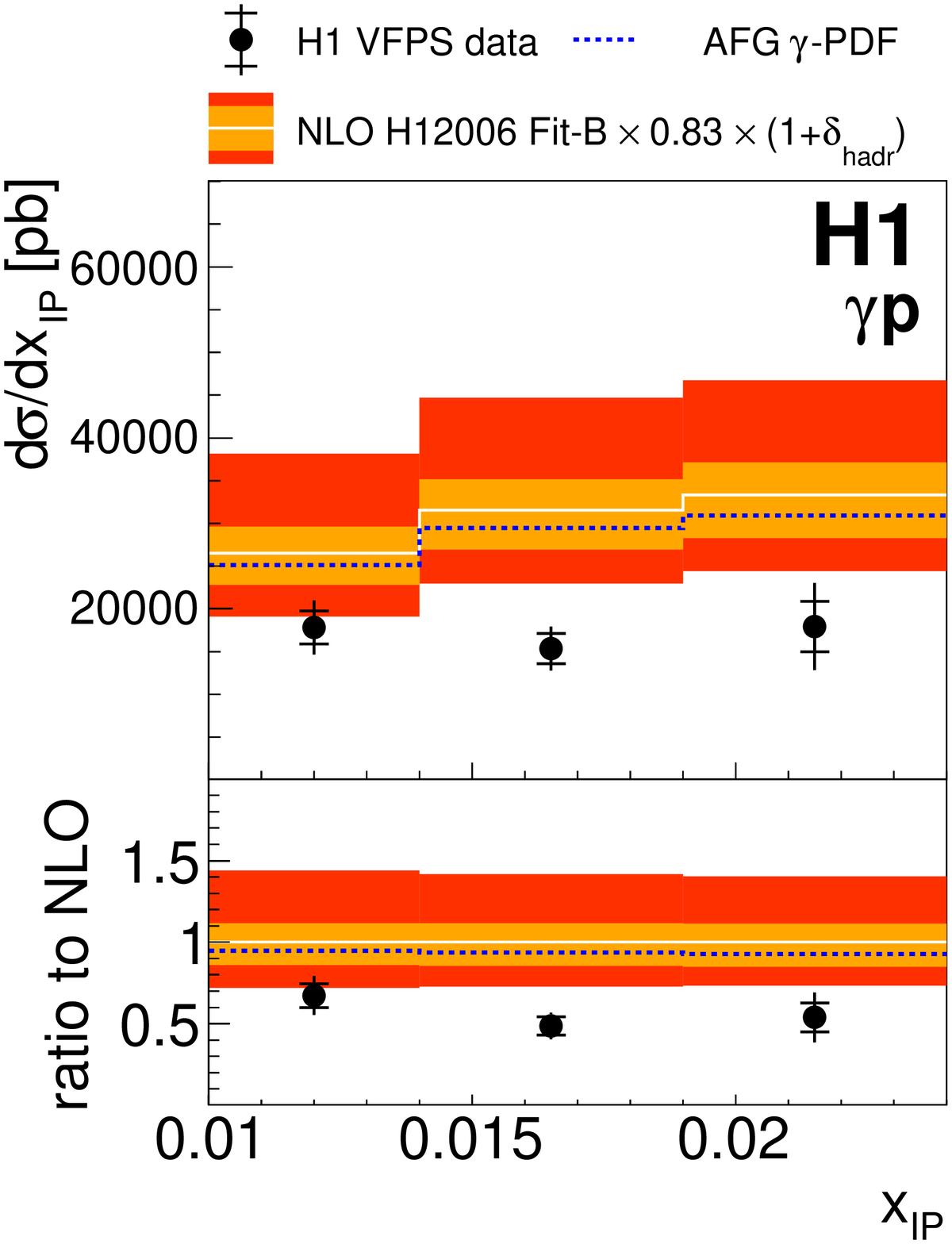}
\includegraph[width=0.4\textwidth]{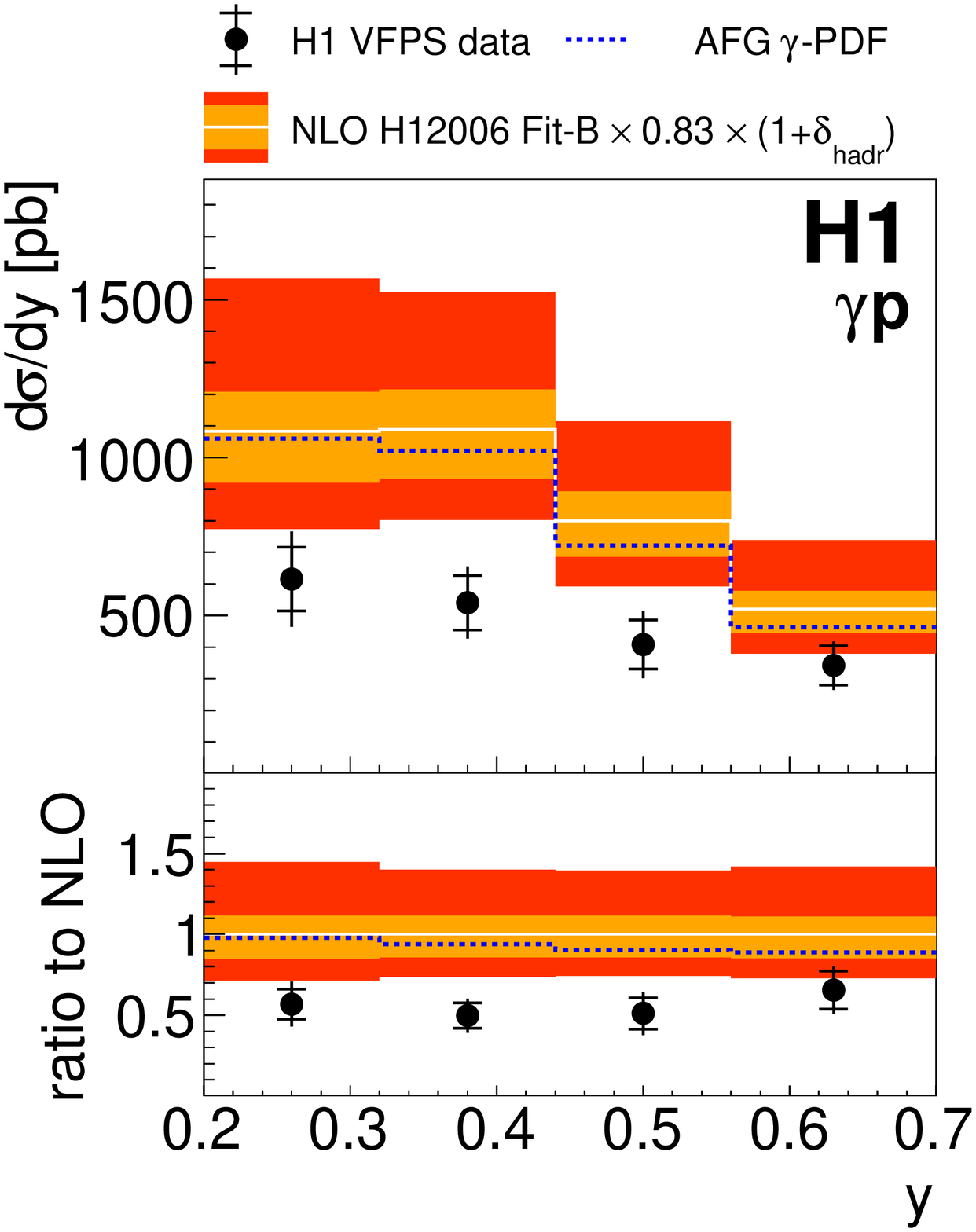}
\includegraph[width=0.4\textwidth]{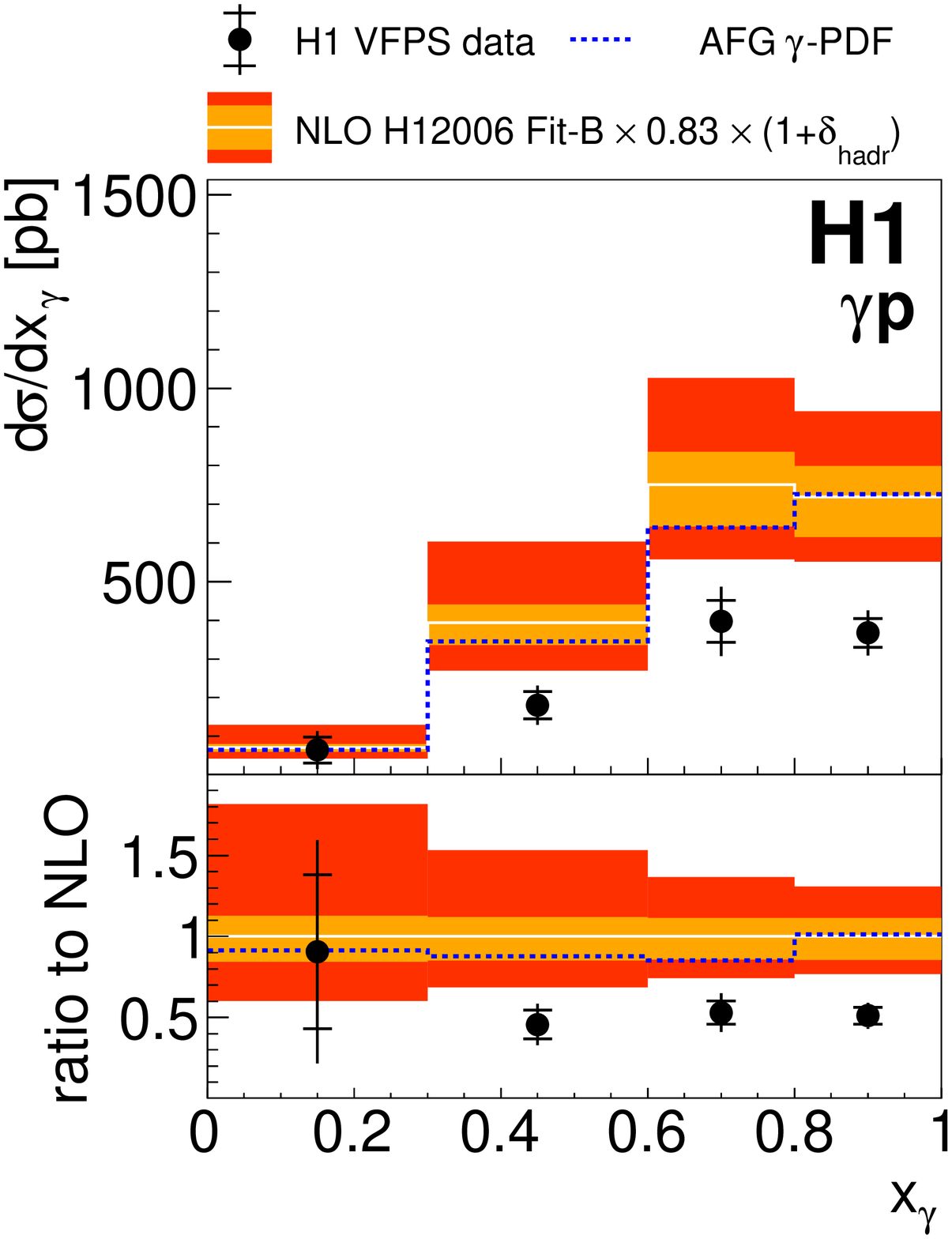}
\end{center}
\caption{Diffractive dijet $ep$ cross sections in the photoproduction
  kinematic range differential in $z_{\pom}$, $x_{\pom}$, $y$
  and $x_{\gamma}$.
  The inner error bars represent the statistical errors.
  The outer error bars indicate the statistical and systematic errors
  added in quadrature.
  The overall normalisation uncertainty of $6\%$ is not shown.
  NLO QCD predictions based on the H12006 Fit-B DPDF set
  and the GRV $\gamma$-PDF set,
  corrected to the level of stable
  hadrons, are shown as a white line. 
  They are scaled by a factor $0.83$ to account for
  contributions from proton-dissociation which are present in the DPDF
  fit but not in the data.
  The inner, light shaded band indicates the size of the DPDF uncertainties
  and hadronisation corrections added in quadrature.
  The outer, dark shaded band indicates the total NLO uncertainty, also
  including scale variations by a factor of $0.5$ to $2$.
  A variant of the NLO calculation using the AFG $\gamma$-PDF set
  is shown as a dashed line.
  For each variable, the cross section is shown in the upper panel,
  whereas the ratio to the NLO prediction is shown in the lower panel.
}
\label{fig_PhpXsec1}
\end{figure}

\begin{figure}[p]
\begin{center}
\includegraph[width=0.4\textwidth]{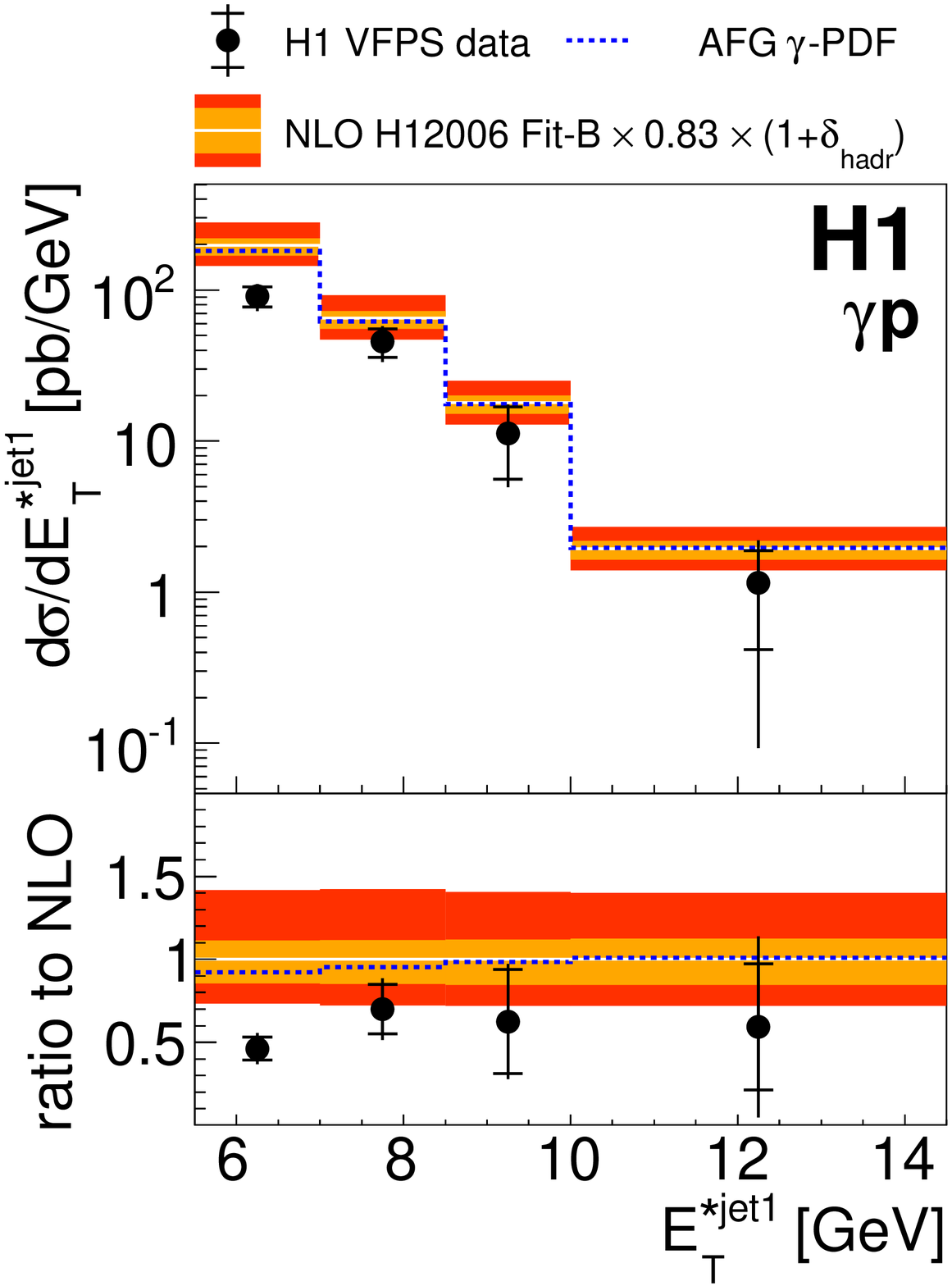}
\includegraph[width=0.4\textwidth]{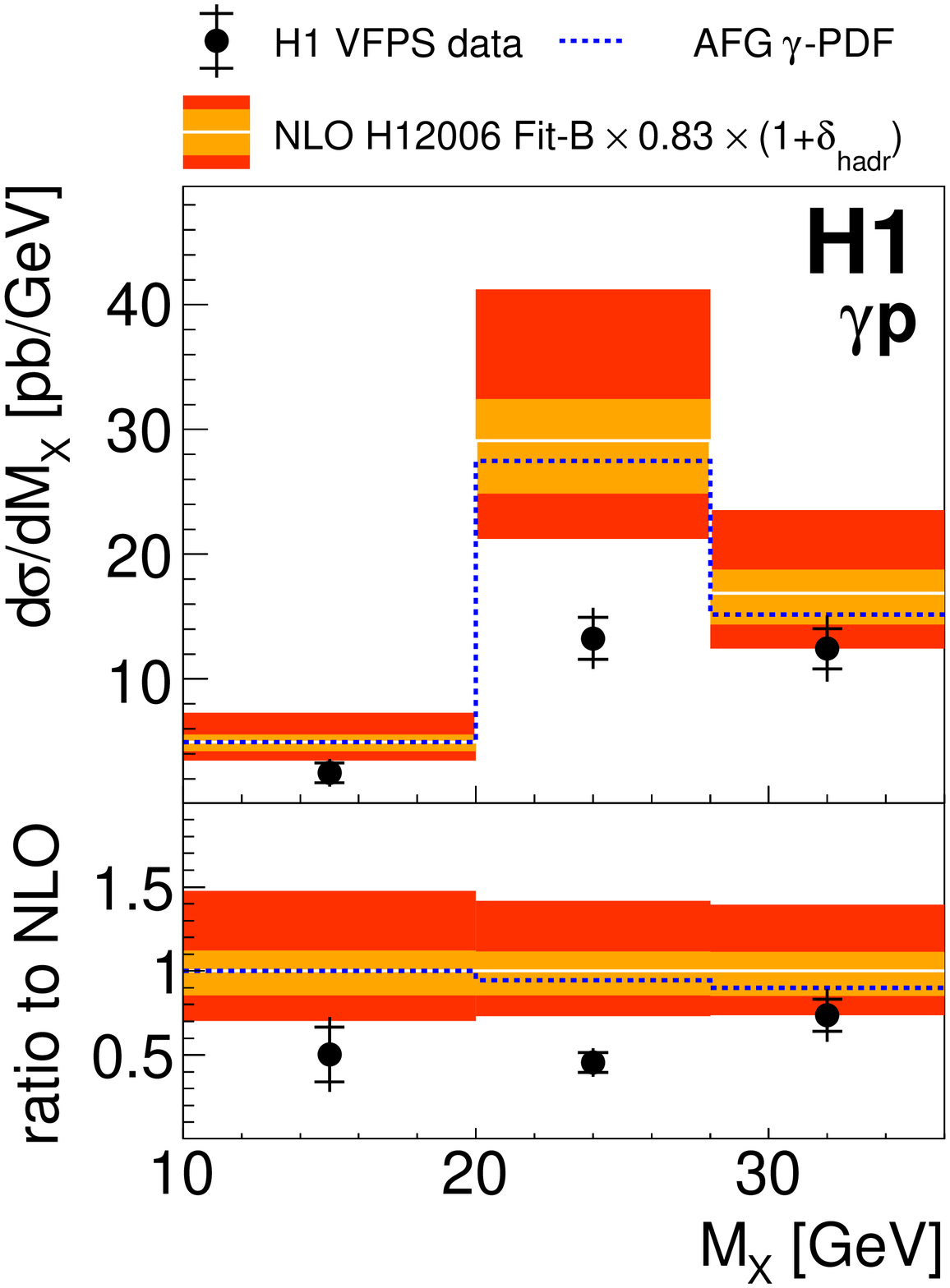}
\includegraph[width=0.4\textwidth]{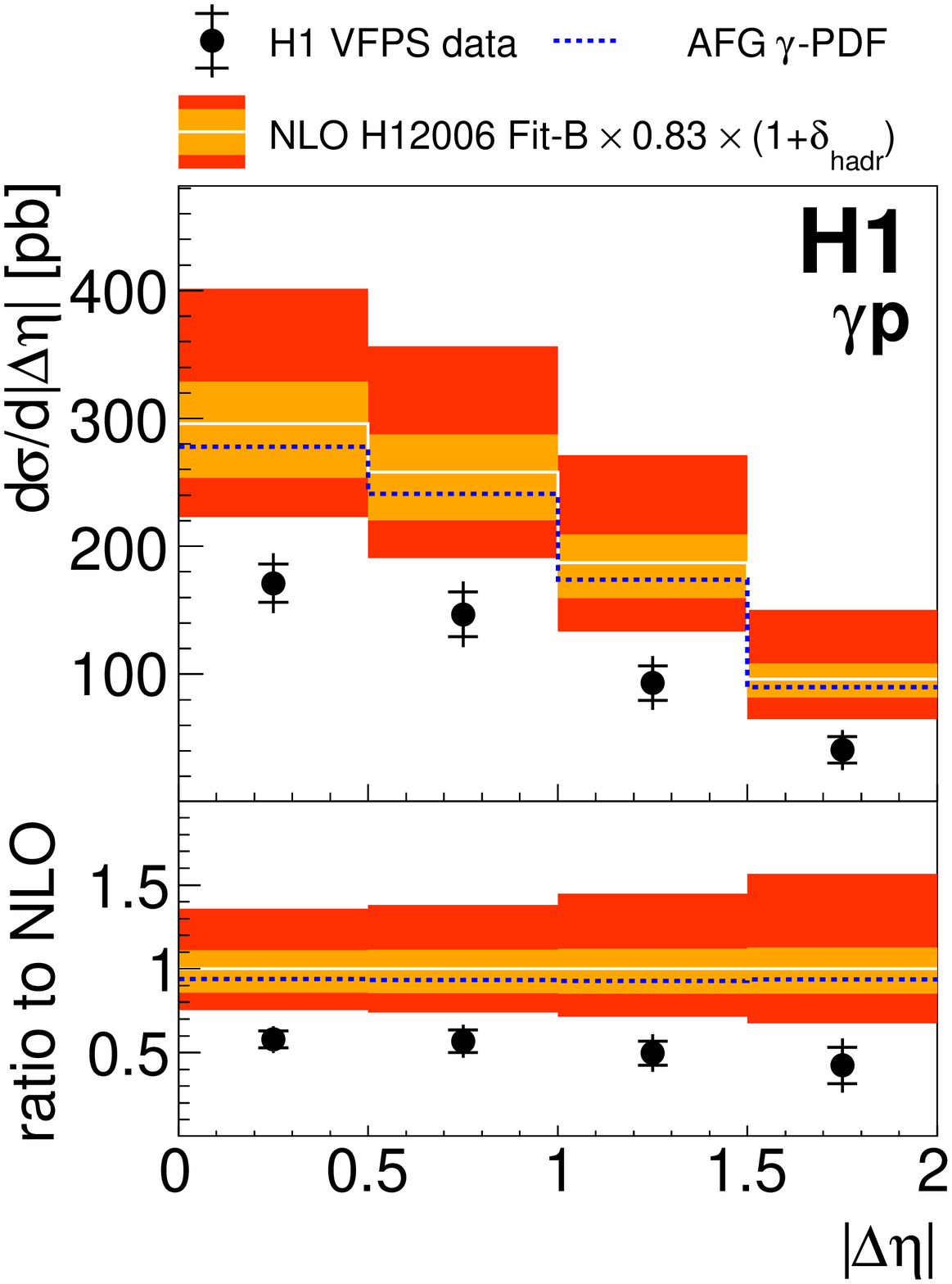}
\includegraph[width=0.4\textwidth]{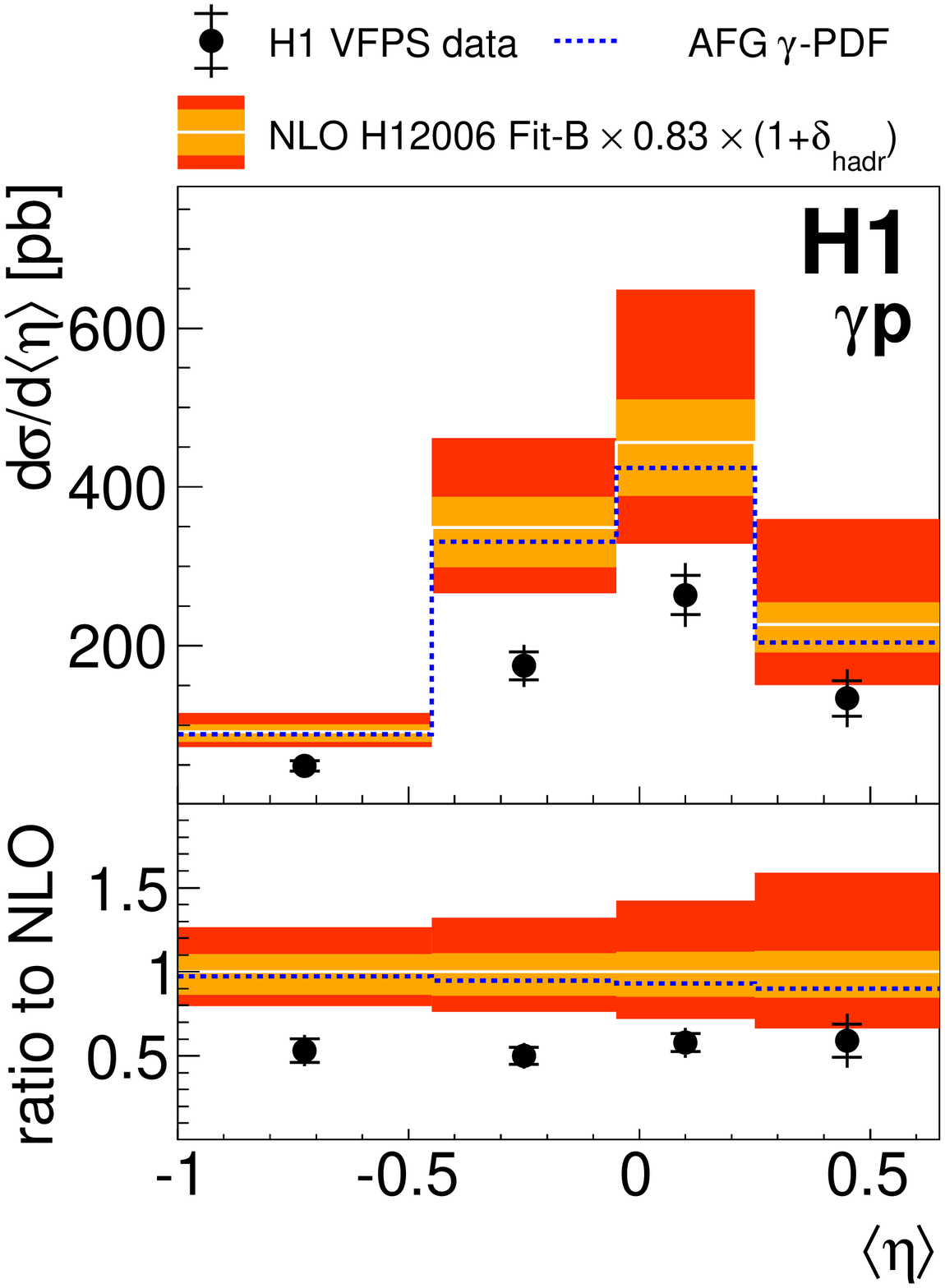}
\end{center}
\caption{Diffractive dijet $ep$ cross sections in the photoproduction
  kinematic range differential in  $E_T^{*\mathrm{jet1}}$,
  $M_X$,
  $|\Delta\eta^{\mathrm{jets}}|$
  and
  $\langle\eta^{\mathrm{jets}}\rangle$.
  The inner error bars represent the statistical errors.
  The outer error bars indicate the statistical and systematic errors
  added in quadrature.
  Further details are given in the caption of figure \ref{fig_PhpXsec1}.
}
\label{fig_PhpXsec2}
\end{figure}

The measured differential cross sections in the $\gamma p$-regime are given in
table~\ref{tab_PhpXsec1} and shown in figure~\ref{fig_PhpXsec1} as a function
of $z_{\pom}^{obs}$, $x_{\pom}$, $y$, $x_{\gamma}^{obs}$ 
together with the NLO QCD calculations.
The differential cross sections for the variables $E_T^{*\mathrm{jet1}}$,
$\langle\eta^{\mathrm{jets}}\rangle$, $\vert\Delta
\eta^{\mathrm{jets}} \vert$
and $M_X$ are given in table~\ref{tab_PhpXsec2} and shown in
figure~\ref{fig_PhpXsec2}.
The relative statistical uncertainties in photoproduction are in
most cases smaller than in the case of deep-inelastic scattering.
%
%
%
The NLO QCD predictions agree well with the measured distributions in
shape but overestimate the dijet cross sections in normalisation, as
already discussed for the integrated cross sections.
%
In particular there is no significant dependence of the data to theory
ratio on  the variables $z_{\pom}^{obs}$, $x_{\gamma}^{obs}$ and
$E_T^{*\mathrm{jet1}}$ 
which are sensitive to the DPDF and to the presence of a diffractive exchange remnant.
These results are in qualitative agreement with previous H1 measurements~\cite{Aaron:2010su,Aktas:2007hn}.
Using the AFG
\cite{Aurenche:2005da} photon PDF as an alternative
the predicted integrated cross section is reduced by $6\%$.
As visible in figures \ref{fig_PhpXsec1} and \ref{fig_PhpXsec2}, the
shapes of the distributions depend only little on the choice of 
the photon PDF.

\subsection{Comparison of Dijet Cross Sections in Diffractive
  Photoproduction and DIS}
\begin{figure}[p]
\begin{center}
\includegraph[width=0.8\textwidth]{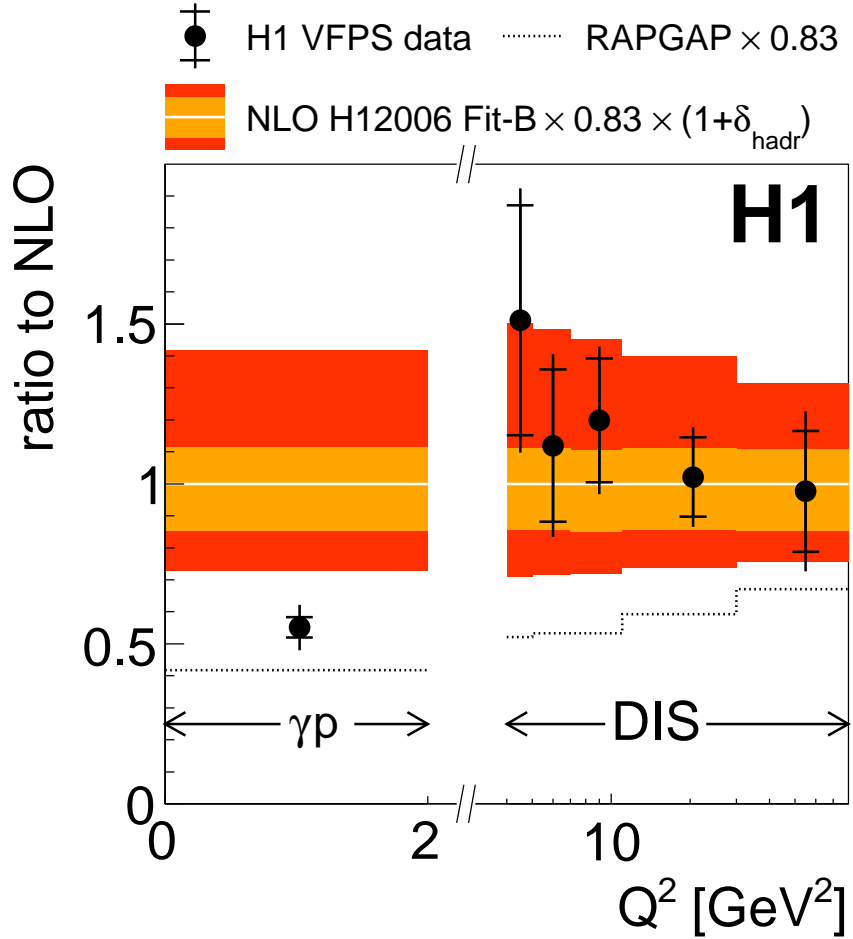}
\end{center}
\caption{Diffractive dijet cross sections in the $\gamma p$- and in the DIS regime
  normalised to the NLO calculation as a function
  of the photon virtuality $Q^2$.
  The inner error bars represent the statistical errors.
  The outer error bars indicate the statistical and systematic errors
  added in quadrature.
  The data points are displayed at the geometrical bin centre.
  The NLO QCD predictions are based on the H12006 Fit-B DPDF set and, in case of photoproduction,  on the GRV $\gamma$-PDF set,
  corrected to the level of stable hadrons.
  They are scaled by a factor $0.83$ to account for
  contributions from proton-dissociation which are present in the DPDF
  fit but not in the data.
  The inner, light shaded band indicates the size of the DPDF uncertainties
  and hadronisation corrections added in quadrature.
  The outer, dark shaded band indicates the total NLO uncertainty, also
  including scale variations by a factor of $0.5$ to $2$.
  Also shown is the ratio of the RAPGAP MC to the NLO prediction.
}
\label{fig_DisPhpQ2}
\end{figure}

The conclusions made in previous sections about the normalisation
problems of the NLO calculations in diffractive photoproduction suffer
from large theoretical uncertainties.
This situation is summarised in figure \ref{fig_DisPhpQ2}, where the
ratio of observed cross section to expectation is shown as a function
of $Q^2$, also including an extra bin for the  cross
section in the photoproduction regime, $0<Q^2<2\,\text{GeV}^2$.
No significant deviation from unity is observed for the suppression
factor\footnote{The suppression factor is defined as a ratio of data
and NLO QCD cross section.}
as a function of $Q^2$ in the DIS regime, whereas
the NLO calculation fails to describe the measurement in
the photoproduction region.
For comparison, also the ratio of the RAPGAP prediction to the NLO
calculation is shown.
RAPGAP is off in normalisation and predicts a shape in $Q^2$ which
differs from the NLO calculation.

\begin{figure}[p]
  \begin{center}
\includegraph[width=0.8\textwidth]{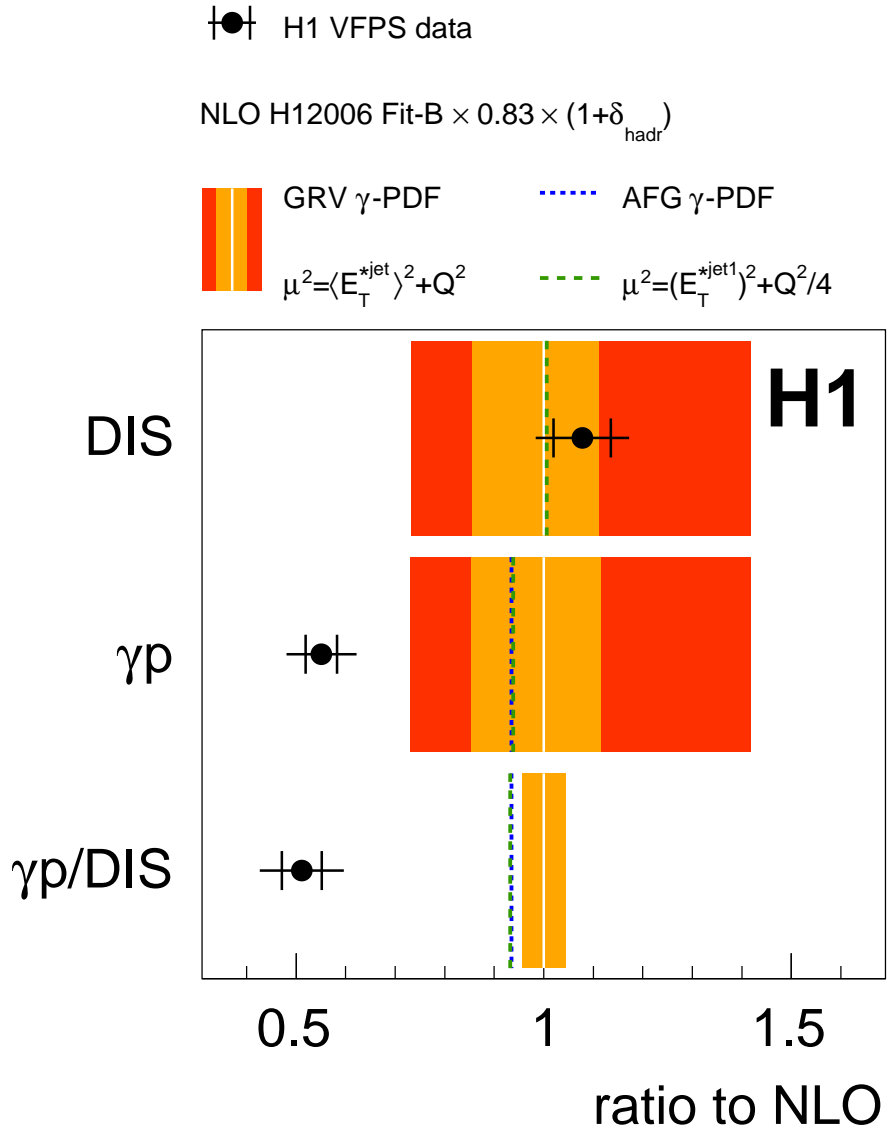} 
\end{center}
\caption{Diffractive dijet DIS and photoproduction cross sections
  normalised to the NLO calculation.
  Also shown is the double ratio of photoproduction to DIS cross
  sections, normalised to the corresponding ratio of NLO predictions.
  The inner error bars represent the statistical errors.
  The outer error bars indicate the statistical and systematic errors
  added in quadrature.
  The NLO QCD predictions are based on the H12006 Fit-B DPDF set and GRV
  $\gamma$-PDF, corrected to the level of stable hadrons.
  They are scaled by a factor $0.83$ to account for
  contributions from proton-dissociation which are present in the DPDF
  fit but not in the data.
  The inner, light shaded band indicates the size of the DPDF uncertainties
  and hadronisation corrections added in quadrature.
  The outer, dark shaded band indicates the total NLO uncertainty, also
  including scale variations by a factor of $0.5$ to $2$.
  Variants of the NLO calculation, normalised to the default NLO prediction,
  are also shown: the effect of using
  the AFG $\gamma$-PDF parametrisation is
  studied in photoproduction. An alternative functional form of the
  scale is studied both in DIS and in photoproduction.
}
\label{fig_DisPhpTotal}
\end{figure}

In a refined method for studying deviations of the NLO QCD predictions
from photoproduction data 
the cross sections measured in the $\gamma p$-regime are divided by the corresponding cross sections in DIS.
In such ratios most of the data systematic uncertainties are reduced,
with the exception of the model uncertainties which are uncorrelated between
$\gamma p$ and DIS.
Similarly, theoretical uncertainties cancel to a large
extent. This is true for the DPDF uncertainties as well as for scale
variations, if the NLO QCD scales are varied simultaneously for
photoproduction and DIS.
The hadronisation corrections, however, are taken to be uncorrelated
between DIS and photoproduction, such that they amount to about
$\sqrt{2}\times 3\%$ in the ratio of the integrated cross section.
The resulting cross section ratios of photoproduction to DIS are
summarised in table \ref{tab_xsecratio}.
\begin{table}[h]
\begin{center}
%
%
%
%
%
\renewcommand{\arraystretch}{1.2}
\setlength{\tabcolsep}{0.1em}
\newcommand{\DATASTATSYSTOT}[6]{$#3$&$\pm#4$ (stat)&$\pm#5$ (syst)& }
\newcommand{\NLO}[4]{#1&#2 (scale)&#3 (DPDF) &#4 (hadr)}
\newcommand{\AFG}[1]{$\,#1$}
\newcommand{\Qzero}[1]{$\,#1$}
\newcommand{\Qone}[1]{$\,#1$}
\newcommand{\Qfour}[1]{$\,#1$}
\newcommand{\RATIODATATHEORY}[3]{#1&#2 (data)&#3 (theory)& }
\newcommand{\RATIOAFG}[1]{$\,#1$}
\newcommand{\RATIOQzero}[1]{$\,#1$}
\newcommand{\RATIOQone}[1]{$\,#1$}
\newcommand{\RATIOQfour}[1]{$\,#1$}
\begin{tabular}{l|rlll} \hline
 & \multicolumn{4}{c}{Ratio of photoproduction to DIS} \\ \hline\hline
Data &
\DATASTATSYSTOT{ }{ }{7.78}{0.60}{1.14}{0.887}
\\
NLO QCD &
\NLO{$15.21$}{${}^{+0.00}_{-0.04}$}{${}^{+0.21}_{-0.10}$}{$\pm0.65$}
\\ &
\AFG{14.22}
 & \multicolumn{3}{l}{\footnotesize$\quad$with AFG $\gamma$PDF} \\ &
\Qfour{14.17}
 & \multicolumn{3}{l}{\footnotesize$\quad$with scale $\mu^2=(E_T^{*\mathrm{jet1}})^2+Q^2/4$}
\\ \hline\hline
Data/NLO &
\RATIODATATHEORY{$0.511$}{$\pm0.085$}{${}^{+0.022}_{-0.021}$}
\\ &
\RATIOAFG{0.547}
 & \multicolumn{3}{l}{\footnotesize$\quad$with AFG $\gamma$PDF} \\ &
\RATIOQfour{0.548}
 & \multicolumn{3}{l}{\footnotesize$\quad$with scale $\mu^2=(E_T^{*\mathrm{jet1}})^2+Q^2/4$} \\ \hline
\end{tabular}
\end{center}
\caption{Ratio of integrated $e^{+}p$ diffractive dijet cross sections for
  $Q^2<2\,\text{GeV}^2$ (photoproduction) to $Q^2>4\,\text{GeV}^2$
  (DIS).
  Listed are the ratios for data and for the NLO calculation including
  two variants.
  The data and NLO uncertainties are indicated.
  The double-ratio of data to NLO and its uncertainties are also given.}
\label{tab_xsecratio}
\end{table}
The double-ratio of photoproduction to DIS, data to NLO, is also given
and shown in figure~\ref{fig_DisPhpTotal}.
Due to the reduced theoretical uncertainty the double ratio deviates
significantly from unity
indicating that factorisation does
not hold in diffractive dijet photoproduction with respect to the same
process in DIS.
This statement is valid within the theoretical framework applied in
this paper and under the assumption that the scale must be varied 
simultaneously for the DIS and $\gamma$p calculations, which leads to
cancellations of the respective uncertainties in the ratio.
Higher order corrections may change this picture.
As an estimate of the possible size of such corrections 
the difference between leading-order and NLO calculations scaled
by $\alpha_s/2$ may be taken, 
which amounts to $5\%$.
When changing the photon PDF from the GRV PDF set to the  AFG PDF set a rise in the
double ratio of $6\%$ is observed.
Using $\mu^2 = {E_T^{*\mathrm{jet1}}}^2+Q^2/4$ as the scale choice leads to an increase of the double ratio by $7\%$.
%
The observed suppression agrees with previous H1 results
\cite{Aaron:2010su,Aktas:2007hn}.
It is worth mentioning that the suppression is now measured at HERA both in
processes with an identified leading proton and in processes with a large
rapidity gap selection, so possible contributions from proton-dissociative
processes alone are excluded as an explanation.

\begin{figure}[p]
\begin{center}
\includegraph[width=0.4\textwidth]{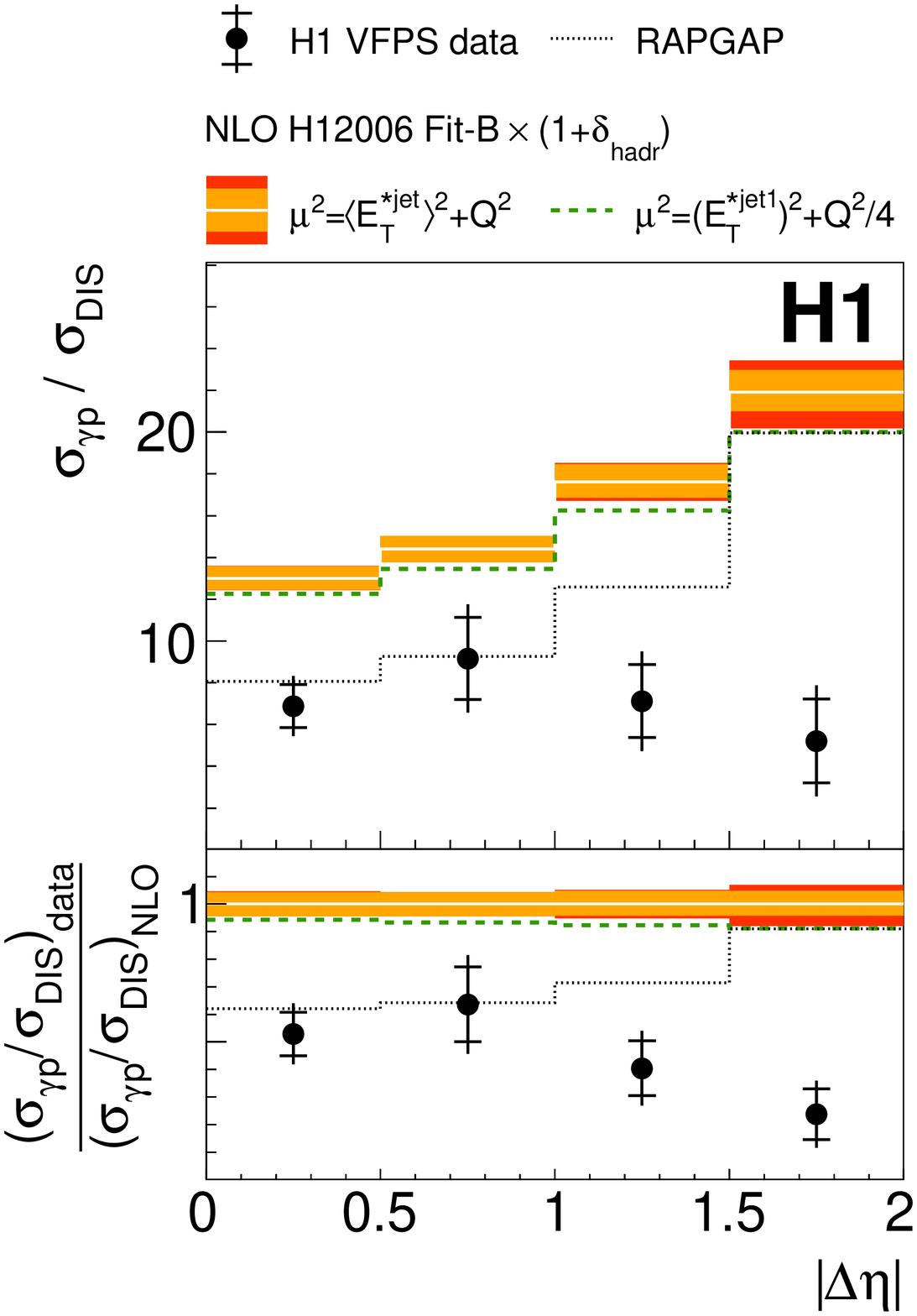}
\includegraph[width=0.4\textwidth]{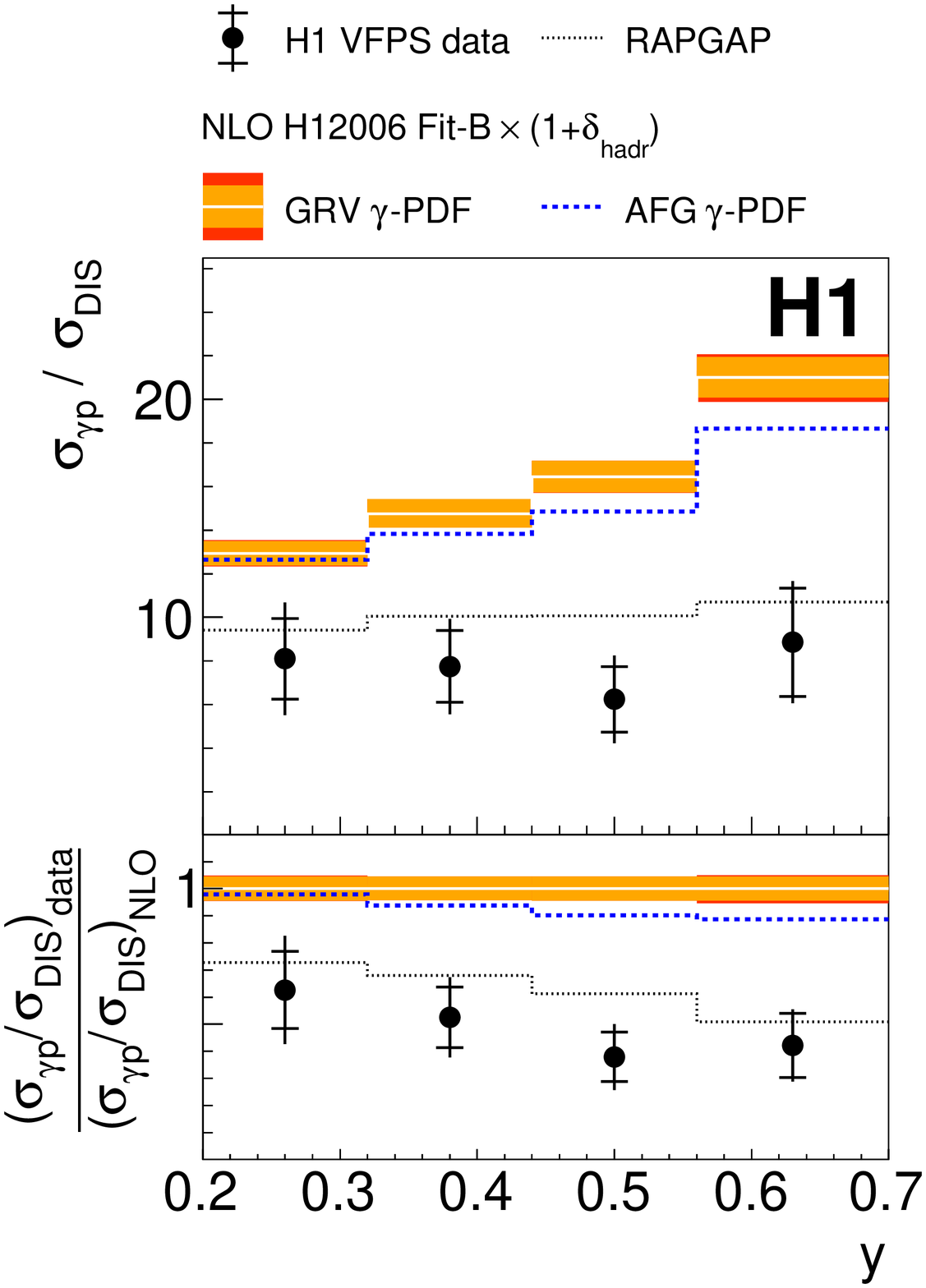}
\end{center}
\caption{Ratios of diffractive dijet photoproduction to DIS cross
  sections differential in $\vert\Delta\eta\vert$ and $y$.
  The inner error bars represent the statistical errors.
  The outer error bars indicate the statistical and systematic errors
  added in quadrature.
  The NLO QCD predictions are based on the H12006 Fit-B DPDF set
  and GRV $\gamma$-PDF, corrected to the level of stable hadrons.
  They are scaled by a factor $0.83$ to account for
  contributions from proton-dissociation which are present in the DPDF
  fit but not in the data.
  The inner, light shaded band indicates the size of the DPDF uncertainties
  and hadronisation corrections added in quadrature.
  The outer, dark shaded band indicates the total NLO uncertainty, also
  including scale variations by a factor of $0.5$ to $2$.
  Variants of the NLO calculation, normalised to the default calculation,
  are also shown. 
  An alternative functional form of the
  scale is studied differential in $\vert\Delta\eta\vert$.
  The effect of using
  the AFG $\gamma$-PDF parametrisation is studied differential in $y$. 
}
\label{fig_DisPhpRatio1}
\end{figure}

\begin{figure}[p]
\begin{center}
\includegraph[width=0.4\textwidth]{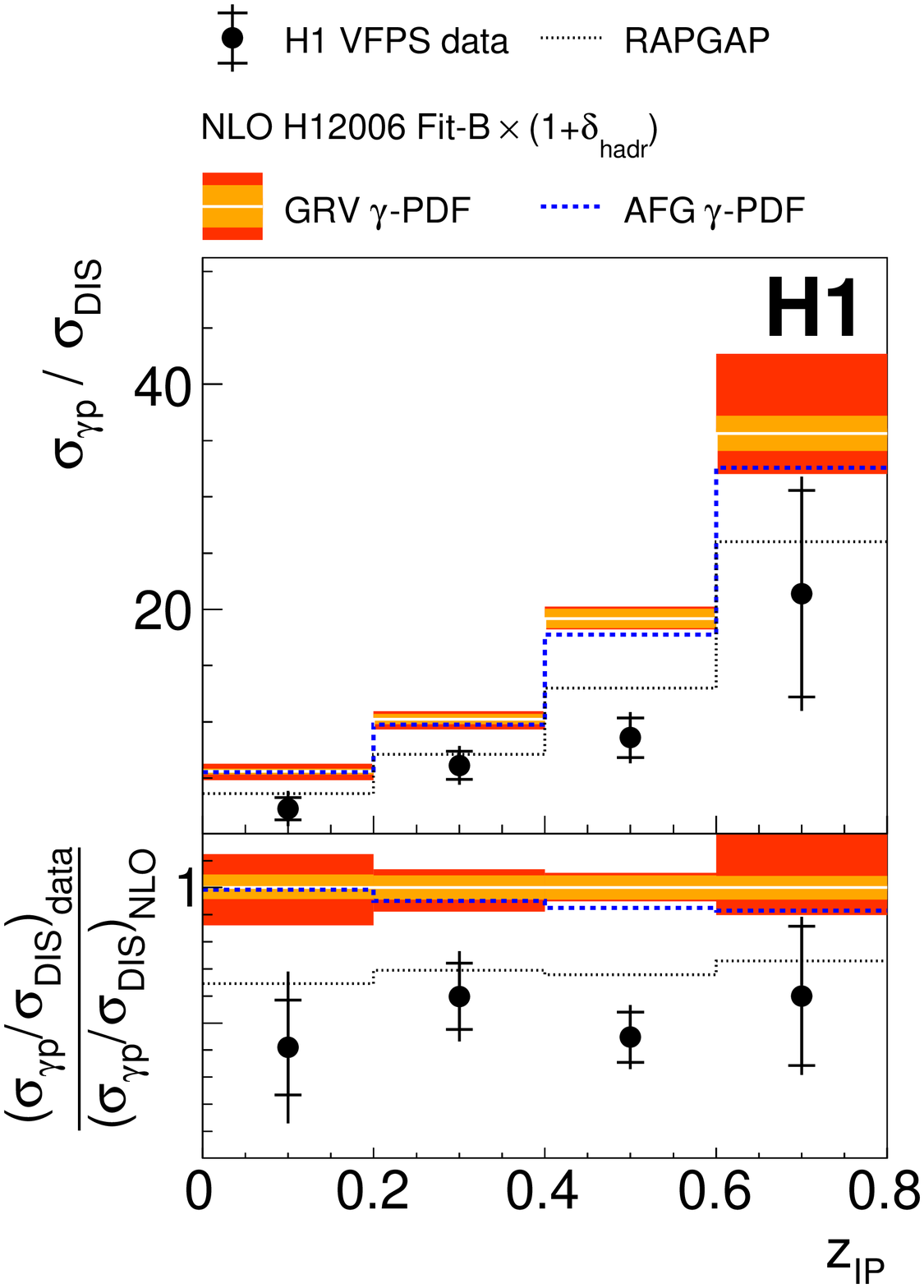}
\includegraph[width=0.4\textwidth]{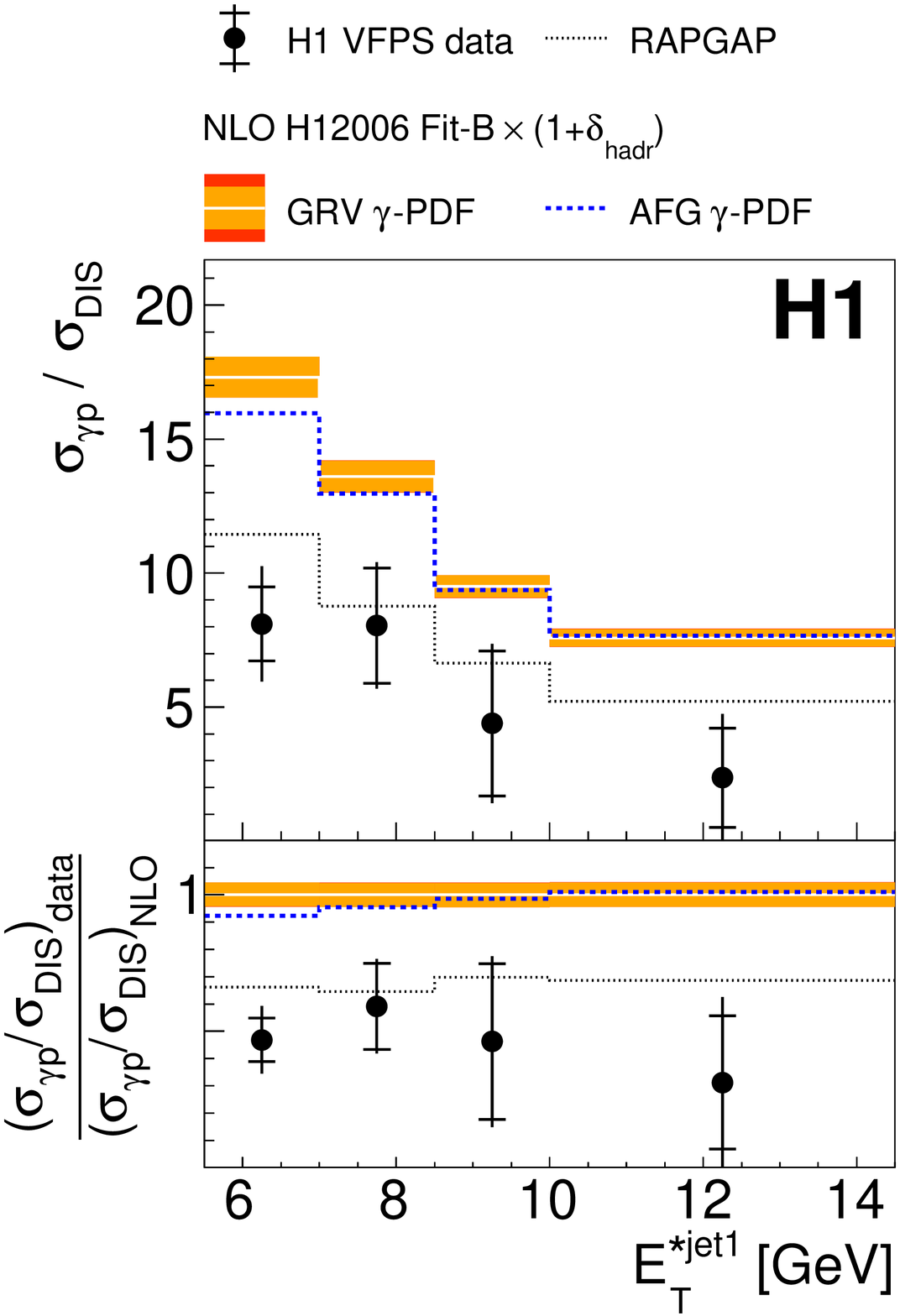}
\end{center}
\caption{Ratios of diffractive dijet photoproduction to DIS cross
  sections differential in $z_\pom$ and $E_T^{*\mathrm{jet1}}$.
  The inner error bars represent the statistical errors.
  The outer error bars indicate the statistical and systematic errors
  added in quadrature.
  The NLO QCD predictions are based on the H12006 Fit-B DPDF set and GRV
  $\gamma$-PDF, corrected to the level of stable hadrons.
  They are scaled by a factor $0.83$ to account for
  contributions from proton-dissociation which are present in the DPDF
  fit but not in the data.
  The inner, light shaded band indicates the size of the DPDF uncertainties
  and hadronisation corrections added in quadrature.
  The outer, dark shaded band indicates the total NLO uncertainty, also
  including scale variations by a factor of $0.5$ to $2$.
  A variant of the NLO calculation using the AFG $\gamma$-PDF
  is shown as a dashed line.
}
\label{fig_DisPhpRatio2}
\end{figure}

Possible shape dependencies of the suppression are studied using cross
section ratios of photoproduction to DIS differential in the variables
$\vert\Delta \eta^{\mathrm{jets}} \vert$, $y$, $z_{\pom}$ and
$E_T^{*\mathrm{jet1}}$, as given in table~\ref{tab_xsecRatio}.
The data ratios as a function of $\vert\Delta\eta^{\mathrm{jets}} \vert$
and $y$ are shown in figure \ref{fig_DisPhpRatio1} together with predictions from
NLO QCD and RAPGAP.
The measured shapes are not described well, but the limited
experimental precision does not allow for strong conclusions to be made.
The ratios as a function of $z_{\pom}$
and $E_T^{*\mathrm{jet1}}$ are shown in figure \ref{fig_DisPhpRatio2}.
%
Within uncertainties the corresponding double ratios are constant
throughout the measured $z_{\pom}$ and $E_T^{*\mathrm{jet1}}$ ranges.

\section{Summary}
Diffractive dijet production is measured in photoproduction and
deep-inelastic scattering in the same kinematic range
$0.2<y<0.7$ and $0.010< x_{\pom} < 0.024$ for jets with
$E_T^{*\mathrm{jet1}} > 5.5\, ~\mbox{GeV}$, $E_T^{*\mathrm{jet2}} >
4.0\,\mbox{GeV}$ and with limits on the photon virtuality $Q^2<2
~\mbox{GeV}^2$  for photoproduction and
$4\,\mbox{GeV}^2<Q^2<80\,\mbox{GeV}^2$ for DIS.
For the leading proton detection, the H1 Very Forward Proton
Spectrometer is used for the first time, such that the diffractive
sample is free of background from low-mass proton dissociative
states.

In DIS, diffractive dijet production is well described 
within the
experimental and theoretical uncertainties
by the NLO calculations based
on the H12006 Fit-B diffractive parton densities of the proton.
Within uncertainties, the QCD factorisation assumptions made for the
NLO calculation are confirmed in this process. 
This result is consistent with previous H1 and ZEUS measurements and the
new data may be used in future DPDF fits.

In photoproduction, next-to-leading order (NLO) calculations based on the
H12006 Fit-B diffractive parton  densities overestimate the measured total cross
sections, thus confirming previous H1 measurements, where the Large
Rapidity Gap method for the identification of diffractive events was used.
The shapes of the differential cross sections are described within the 
experimental and theoretical uncertainties.
There is no hint of dependence of the observed suppression
on the variable $x_{\gamma}^{obs}$.
%

In order to profit from cancellations of theoretical
uncertainties,  ratios of photoproduction to DIS cross sections and
double ratios of data to NLO are analysed.
Integrated over the analysis phase space the double ratio is found to
be $0.51\pm0.09$.
Following this, within the theoretical framework based on 
diffractive parton densities, 
factorisation is broken in diffractive dijet photoproduction.
This observation is in agreement with previous H1 measurements, where
complementary experimental methods have been used.
Contributions from proton dissociative processes present in the
previous analyses but absent here are ruled out as a cause of the
observed suppression.
%
The differential measurements of cross sections and
cross section ratios in DIS and photoproduction provide stringent
tests of the theory both in normalisation and in shape.

\section*{Acknowledgements}

We are grateful to the HERA machine group whose outstanding
efforts have made this experiment possible. 
We thank the engineers and technicians for their work in constructing and
maintaining the H1 detector, our funding agencies for 
financial support, the
DESY technical staff for continual assistance
and the DESY directorate for support and for the
hospitality which they extend to the non DESY 
members of the collaboration.
We would like to give credit to all partners contributing to the EGI 
computing infrastructure for their support for the H1 Collaboration.
We express our gratitude to M. Ryskin for helpful discussions.

\bibliographystyle{utcaps_new}
\begin{flushleft}
\providecommand{\href}[2]{#2}\begingroup\raggedright\endgroup
\end{flushleft}

\newpage

\begin{sidewaystable}[p]
\begin{center}
\footnotesize
%
%
%
%
%
\setlength{\tabcolsep}{0.3em}
\newcommand{\DATASTATSYSREL}[9]{$#1$&$#2$&$#3$&$#4$&#5&$#6$&#7&$#9$&$#8$\\}
\begin{tabular}{|c|c|r|c|l|c|l|r|r|}
\hline
\multicolumn{2}{|c|}{integrated}&\multicolumn{1}{c|}{$\sigma$}&
\multicolumn{1}{c|}{$\delta_{\text{stat}}$}&
\multicolumn{1}{c|}{$\rho_{\text{stat}}$}&
\multicolumn{1}{c|}{$\delta_{\text{syst}}$}&
\multicolumn{1}{c|}{$\rho_{\text{syst}}$}&
\multicolumn{1}{c|}{$1+\delta_{\text{rad}}$}&
\multicolumn{1}{c|}{$1+\delta_{\text{hadr}}$} \\
\multicolumn{2}{|c|}{cross section}&\multicolumn{1}{c|}{[pb]}&[$\%$]&\multicolumn{1}{c|}{[$\%$]}&
[$\%$]&\multicolumn{1}{c|}{[$\%$]}& & \\
\hline
\DATASTATSYSREL{ }{ }{30.5}{5.4}{}{9.0}{}{1.022}{0.999}
\hline
\hline
$z_\pom$&bin&\multicolumn{1}{c|}{$d\sigma/dz_\pom$}&
\multicolumn{1}{c|}{$\delta_{\text{stat}}$}&
\multicolumn{1}{c|}{$\rho_{\text{stat}}$}&
\multicolumn{1}{c|}{$\delta_{\text{syst}}$}&
\multicolumn{1}{c|}{$\rho_{\text{syst}}$}&
\multicolumn{1}{c|}{$1+\delta_{\text{rad}}$}&
\multicolumn{1}{c|}{$1+\delta_{\text{hadr}}$} \\
 & &\multicolumn{1}{c|}{[pb]}&[$\%$]&\multicolumn{1}{c|}{[$\%$]}&
[$\%$]&\multicolumn{1}{c|}{[$\%$]}& & \\
\hline
\DATASTATSYSREL{0\div 0.2}{1}{32.1}{20.9}{}{16.6}{}{1.084}{0.916}
\DATASTATSYSREL{0.2\div 0.4}{2}{59.8}{12.4}{$\rho_{12}=-17$}{10.6}{$\rho_{12}=74$}{1.054}{1.012}
\DATASTATSYSREL{0.4\div 0.6}{3}{48.0}{14.9}{$\rho_{13}=-4$ $\rho_{23}=-18$}{9.9}{$\rho_{13}=33$ $\rho_{23}=79$}{0.996}{1.017}
\DATASTATSYSREL{0.6\div 0.8}{4}{13.9}{39.0}{$\rho_{14}=4$ $\rho_{24}=1$ $\rho_{34}=-37$}{16.6}{$\rho_{14}=-12$ $\rho_{24}=28$ $\rho_{34}=17$}{0.910}{1.028}
\hline
\hline
$x_\pom$&bin&\multicolumn{1}{c|}{$d\sigma/dx_\pom$}&
\multicolumn{1}{c|}{$\delta_{\text{stat}}$}&
\multicolumn{1}{c|}{$\rho_{\text{stat}}$}&
\multicolumn{1}{c|}{$\delta_{\text{syst}}$}&
\multicolumn{1}{c|}{$\rho_{\text{syst}}$}&
\multicolumn{1}{c|}{$1+\delta_{\text{rad}}$}&
\multicolumn{1}{c|}{$1+\delta_{\text{hadr}}$} \\
 & &\multicolumn{1}{c|}{[pb]}&[$\%$]&\multicolumn{1}{c|}{[$\%$]}&
[$\%$]&\multicolumn{1}{c|}{[$\%$]}& & \\
\hline
\DATASTATSYSREL{0.01\div 0.014}{1}{2250}{14.3}{}{20.1}{}{1.058}{0.998}
\DATASTATSYSREL{0.014\div 0.019}{2}{2210}{12.8}{$\rho_{12}=-14$}{14.4}{$\rho_{12}=-33$}{1.014}{1.003}
\DATASTATSYSREL{0.019\div 0.024}{3}{2290}{12.5}{$\rho_{13}=4$ $\rho_{23}=-18$}{12.0}{$\rho_{13}=25$ $\rho_{23}=41$}{1.006}{0.997}
\hline
\hline
$y$&bin&\multicolumn{1}{c|}{$d\sigma/dy$}&
\multicolumn{1}{c|}{$\delta_{\text{stat}}$}&
\multicolumn{1}{c|}{$\rho_{\text{stat}}$}&
\multicolumn{1}{c|}{$\delta_{\text{syst}}$}&
\multicolumn{1}{c|}{$\rho_{\text{syst}}$}&
\multicolumn{1}{c|}{$1+\delta_{\text{rad}}$}&
\multicolumn{1}{c|}{$1+\delta_{\text{hadr}}$} \\
 & &\multicolumn{1}{c|}{[pb]}&[$\%$]&\multicolumn{1}{c|}{[$\%$]}&
[$\%$]&\multicolumn{1}{c|}{[$\%$]}& & \\
\hline
\DATASTATSYSREL{0.2\div 0.32}{1}{76}{15.7}{}{13.4}{}{0.992}{1.060}
\DATASTATSYSREL{0.32\div 0.44}{2}{69.7}{14.1}{$\rho_{12}=-12$}{11.7}{$\rho_{12}=86$}{1.002}{0.975}
\DATASTATSYSREL{0.44\div 0.56}{3}{65.4}{14.7}{$\rho_{13}=0$ $\rho_{23}=-12$}{10.7}{$\rho_{13}=63$ $\rho_{23}=73$}{1.056}{0.992}
\DATASTATSYSREL{0.56\div 0.7}{4}{38.6}{21.4}{$\rho_{14}=-1$ $\rho_{24}=1$ $\rho_{34}=-17$}{10.4}{$\rho_{14}=39$ $\rho_{24}=59$ $\rho_{34}=59$}{1.084}{0.948}
\hline
\hline
$Q^2$&bin&\multicolumn{1}{c|}{$d\sigma/dQ^2$}&
\multicolumn{1}{c|}{$\delta_{\text{stat}}$}&
\multicolumn{1}{c|}{$\rho_{\text{stat}}$}&
\multicolumn{1}{c|}{$\delta_{\text{syst}}$}&
\multicolumn{1}{c|}{$\rho_{\text{syst}}$}&
\multicolumn{1}{c|}{$1+\delta_{\text{rad}}$}&
\multicolumn{1}{c|}{$1+\delta_{\text{hadr}}$} \\
{}[GeV${}^2$]& &\multicolumn{1}{c|}{[pb/GeV${}^2$]}&[$\%$]&\multicolumn{1}{c|}{[$\%$]}&
[$\%$]&\multicolumn{1}{c|}{[$\%$]}& & \\
\hline
\DATASTATSYSREL{4\div 5}{1}{4.83}{23.8}{}{14.7}{}{1.020}{0.982}
\DATASTATSYSREL{5\div 7}{2}{2.55}{21.3}{$\rho_{12}=-17$}{15.3}{$\rho_{12}=36$}{1.020}{1.002}
\DATASTATSYSREL{7\div 11}{3}{1.66}{16.1}{$\rho_{13}=3$ $\rho_{23}=-11$}{12.1}{$\rho_{13}=80$ $\rho_{23}=63$}{1.028}{0.974}
\DATASTATSYSREL{11\div 30}{4}{0.520}{12.1}{$\rho_{14}=1$ $\rho_{24}=3$ $\rho_{34}=-4$}{11.1}{$\rho_{14}=39$ $\rho_{24}=66$ $\rho_{34}=71$}{1.034}{1.019}
\DATASTATSYSREL{30\div 80}{5}{0.104}{19.4}{$\rho_{15}=1$ $\rho_{25}=2$ $\rho_{35}=3$ $\rho_{45}=-1$}{17.6}{$\rho_{15}=-7$ $\rho_{25}=23$ $\rho_{35}=27$ $\rho_{45}=78$}{1.013}{1.036}
\hline
\end{tabular}
\end{center}
\caption{Integrated cross section and bin averaged hadron level differential
  cross sections as a function of the
  variables $z_\pom$, $x_\pom$, $y$ and $Q^2$ for diffractive dijet
  DIS in the phase space detailed in table \ref{tab_cuts}.
  For each data point, the statistical ($\delta_{\text{stat}}$) and
  systematic ($\delta_{\text{syst}}$) uncertainties and the
  corresponding correlation
  coefficients ($\rho_{\text{stat}}$, $\rho_{\text{syst}}$) are given.
  The hadronisation correction factors
  ($1+\delta_{\text{hadr}}$) applied to the NLO calculations and the
  radiative corrections ($1+\delta_{\text{rad}}$) are also listed.
 The overall normalisation uncertainty of $6\%$ is not included in the table.}
\label{tab_DisXsec1}
\end{sidewaystable}

\begin{sidewaystable}[p]
\begin{center}
\footnotesize
%
%
%
%
%
\setlength{\tabcolsep}{0.3em}
\newcommand{\DATASTATSYSREL}[9]{$#1$&$#2$&$#3$&$#4$&#5&$#6$&#7&$#9$&$#8$\\}
\begin{tabular}{|c|c|r|c|l|c|l|r|r|}
\hline
$E_T^{*\mathrm{jet1}}$&bin&\multicolumn{1}{c|}{$d\sigma/dE_T^{*\mathrm{jet1}}$}&
\multicolumn{1}{c|}{$\delta_{\text{stat}}$}&
\multicolumn{1}{c|}{$\rho_{\text{stat}}$}&
\multicolumn{1}{c|}{$\delta_{\text{syst}}$}&
\multicolumn{1}{c|}{$\rho_{\text{syst}}$}&
\multicolumn{1}{c|}{$1+\delta_{\text{rad}}$}&
\multicolumn{1}{c|}{$1+\delta_{\text{hadr}}$} \\
{}[GeV]& &\multicolumn{1}{c|}{[pb/GeV]}&[$\%$]&\multicolumn{1}{c|}{[$\%$]}&
[$\%$]&\multicolumn{1}{c|}{[$\%$]}& & \\
\hline
\DATASTATSYSREL{5.5\div 7}{1}{11.24}{8.3}{}{11.9}{}{1.006}{0.999}
\DATASTATSYSREL{7\div 8.5}{2}{5.66}{16.4}{$\rho_{12}=-4$}{12.6}{$\rho_{12}=-25$}{1.034}{0.986}
\DATASTATSYSREL{8.5\div 10}{3}{2.55}{36.0}{$\rho_{13}=-22$ $\rho_{23}=-21$}{29.8}{$\rho_{13}=-70$ $\rho_{23}=80$}{1.112}{1.050}
\DATASTATSYSREL{10\div 14.5}{4}{0.485}{45.2}{$\rho_{14}=10$ $\rho_{24}=-14$ $\rho_{34}=-39$}{15.0}{$\rho_{14}=-30$ $\rho_{24}=38$ $\rho_{34}=42$}{0.976}{0.961}
\hline
\hline
$M_X$&bin&\multicolumn{1}{c|}{$d\sigma/dM_X$}&
\multicolumn{1}{c|}{$\delta_{\text{stat}}$}&
\multicolumn{1}{c|}{$\rho_{\text{stat}}$}&
\multicolumn{1}{c|}{$\delta_{\text{syst}}$}&
\multicolumn{1}{c|}{$\rho_{\text{syst}}$}&
\multicolumn{1}{c|}{$1+\delta_{\text{rad}}$}&
\multicolumn{1}{c|}{$1+\delta_{\text{hadr}}$} \\
{}[GeV]& &\multicolumn{1}{c|}{[pb/GeV]}&[$\%$]&\multicolumn{1}{c|}{[$\%$]}&
[$\%$]&\multicolumn{1}{c|}{[$\%$]}& & \\
\hline
\DATASTATSYSREL{10\div 20}{1}{0.20}{61.5}{}{120.1}{}{0.977}{1.024}
\DATASTATSYSREL{20\div 28}{2}{2.06}{9.5}{$\rho_{12}=-23$}{10.6}{$\rho_{12}=76$}{1.021}{1.026}
\DATASTATSYSREL{28\div 36}{3}{1.43}{12.5}{$\rho_{13}=3$ $\rho_{23}=-18$}{12.0}{$\rho_{13}=-67$ $\rho_{23}=-16$}{1.046}{0.952}
\hline
\hline
$\vert\Delta\eta^{\text{jets}}\vert$&bin&\multicolumn{1}{c|}{$d\sigma/d\vert\Delta\eta^{\text{jets}}\vert$}&
\multicolumn{1}{c|}{$\delta_{\text{stat}}$}&
\multicolumn{1}{c|}{$\rho_{\text{stat}}$}&
\multicolumn{1}{c|}{$\delta_{\text{syst}}$}&
\multicolumn{1}{c|}{$\rho_{\text{syst}}$}&
\multicolumn{1}{c|}{$1+\delta_{\text{rad}}$}&
\multicolumn{1}{c|}{$1+\delta_{\text{hadr}}$} \\
 & &\multicolumn{1}{c|}{[pb]}&[$\%$]&\multicolumn{1}{c|}{[$\%$]}&
[$\%$]&\multicolumn{1}{c|}{[$\%$]}& & \\
\hline
\DATASTATSYSREL{0\div 0.5}{1}{24.8}{12.1}{}{10.8}{}{1.004}{1.015}
\DATASTATSYSREL{0.5\div 1}{2}{16.0}{17.6}{$\rho_{12}=-16$}{12.7}{$\rho_{12}=81$}{1.046}{1.002}
\DATASTATSYSREL{1\div 1.5}{3}{13.1}{19.9}{$\rho_{13}=4$ $\rho_{23}=-16$}{13.2}{$\rho_{13}=96$ $\rho_{23}=84$}{1.030}{0.968}
\DATASTATSYSREL{1.5\div 2}{4}{7.8}{28.9}{$\rho_{14}=0$ $\rho_{24}=5$ $\rho_{34}=-14$}{18.4}{$\rho_{14}=38$ $\rho_{24}=70$ $\rho_{34}=53$}{1.030}{0.966}
\hline
\hline
$\langle\eta^{\text{jets}}\rangle$&bin&\multicolumn{1}{c|}{$d\sigma/d\langle\eta^{\text{jets}}\rangle$}&
\multicolumn{1}{c|}{$\delta_{\text{stat}}$}&
\multicolumn{1}{c|}{$\rho_{\text{stat}}$}&
\multicolumn{1}{c|}{$\delta_{\text{syst}}$}&
\multicolumn{1}{c|}{$\rho_{\text{syst}}$}&
\multicolumn{1}{c|}{$1+\delta_{\text{rad}}$}&
\multicolumn{1}{c|}{$1+\delta_{\text{hadr}}$} \\
 & &\multicolumn{1}{c|}{[pb]}&[$\%$]&\multicolumn{1}{c|}{[$\%$]}&
[$\%$]&\multicolumn{1}{c|}{[$\%$]}& & \\
\hline
\DATASTATSYSREL{-1\div -0.45}{1}{10.3}{21.1}{}{11.1}{}{0.905}{1.011}
\DATASTATSYSREL{-0.45\div -0.05}{2}{30.1}{11.4}{$\rho_{12}=-13$}{10.6}{$\rho_{12}=42$}{1.011}{1.021}
\DATASTATSYSREL{-0.05\div 0.25}{3}{30.0}{14.9}{$\rho_{13}=1$ $\rho_{23}=-14$}{11.5}{$\rho_{13}=37$ $\rho_{23}=81$}{1.056}{0.994}
\DATASTATSYSREL{0.25\div 0.65}{4}{9.2}{32.3}{$\rho_{14}=-2$ $\rho_{24}=1$ $\rho_{34}=-16$}{20.3}{$\rho_{14}=18$ $\rho_{24}=63$ $\rho_{34}=71$}{1.171}{0.921}
\hline
\end{tabular}
\end{center}
\caption{Bin averaged hadron level differential cross sections for
  diffractive dijet DIS as a function of the variables
  $E_T^{*\text{jet1}}$, $M_X$, $\vert\Delta\eta^{\text{jets}}\vert$
  and $\langle\eta^{\text{jets}}\rangle$ in the phase space detailed
  in table \ref{tab_cuts}.
  For each data point, the statistical ($\delta_{\text{stat}}$) and
  systematic ($\delta_{\text{syst}}$) uncertainties and the
  corresponding correlation
  coefficients ($\rho_{\text{stat}}$, $\rho_{\text{syst}}$) are given.
  The hadronisation correction factors
  ($1+\delta_{\text{hadr}}$) applied to the NLO calculations and the
  the radiative corrections ($1+\delta_{\text{rad}}$) are also listed.
 The overall normalisation uncertainty of $6\%$ is not included in the table.}
\label{tab_DisXsec2}
\end{sidewaystable}

\begin{sidewaystable}[p]
\begin{center}
\footnotesize
%
%
%
%
%
\setlength{\tabcolsep}{0.3em}
\newcommand{\DATASTATSYSREL}[9]{$#1$&$#2$&$#3$&$#4$&#5&$#6$&#7&$#8$\\}
\begin{tabular}{|c|c|r|c|l|c|l|r|}
\hline
\multicolumn{2}{|c|}{integrated}&\multicolumn{1}{c|}{$\sigma$}&
\multicolumn{1}{c|}{$\delta_{\text{stat}}$}&
\multicolumn{1}{c|}{$\rho_{\text{stat}}$}&
\multicolumn{1}{c|}{$\delta_{\text{syst}}$}&
\multicolumn{1}{c|}{$\rho_{\text{syst}}$}&
\multicolumn{1}{c|}{$1+\delta_{\text{hadr}}$} \\
\multicolumn{2}{|c|}{cross section}&\multicolumn{1}{c|}{[pb]}&[$\%$]&[$\%$]&
[$\%$]&[$\%$]& \\
\hline
\DATASTATSYSREL{ }{ }{237}{5.7}{}{13.0}{}{0.906}{}
\hline
\hline
$z_\pom$&bin&\multicolumn{1}{c|}{$d\sigma/dz_\pom$}&
\multicolumn{1}{c|}{$\delta_{\text{stat}}$}&
\multicolumn{1}{c|}{$\rho_{\text{stat}}$}&
\multicolumn{1}{c|}{$\delta_{\text{syst}}$}&
\multicolumn{1}{c|}{$\rho_{\text{syst}}$}&
\multicolumn{1}{c|}{$1+\delta_{\text{hadr}}$} \\
 & &\multicolumn{1}{c|}{[pb]}&[$\%$]&[$\%$]&
[$\%$]&[$\%$]& \\
\hline
\DATASTATSYSREL{0\div 0.2}{1}{73}{37.2}{}{47.0}{}{0.754}{}
\DATASTATSYSREL{0.2\div 0.4}{2}{366}{16.1}{$\rho_{12}=-19$}{17.3}{$\rho_{12}=65$}{0.833}{}
\DATASTATSYSREL{0.4\div 0.6}{3}{413}{14.3}{$\rho_{13}=18$ $\rho_{23}=-33$}{16.1}{$\rho_{13}=32$ $\rho_{23}=83$}{0.928}{}
\DATASTATSYSREL{0.6\div 0.8}{4}{298}{17.9}{$\rho_{14}=-3$ $\rho_{24}=22$ $\rho_{34}=-24$}{18.3}{$\rho_{14}=-16$ $\rho_{24}=41$ $\rho_{34}=81$}{1.017}{}
\hline
\hline
$x_\pom$&bin&\multicolumn{1}{c|}{$d\sigma/dx_\pom$}&
\multicolumn{1}{c|}{$\delta_{\text{stat}}$}&
\multicolumn{1}{c|}{$\rho_{\text{stat}}$}&
\multicolumn{1}{c|}{$\delta_{\text{syst}}$}&
\multicolumn{1}{c|}{$\rho_{\text{syst}}$}&
\multicolumn{1}{c|}{$1+\delta_{\text{hadr}}$} \\
 & &\multicolumn{1}{c|}{[pb]}&[$\%$]&\multicolumn{1}{c|}{[$\%$]}&
[$\%$]&\multicolumn{1}{c|}{[$\%$]}& \\
\hline
\DATASTATSYSREL{0.01\div 0.014}{1}{17800}{10.8}{}{15.4}{}{0.933}{}
\DATASTATSYSREL{0.014\div 0.019}{2}{15300}{11.5}{$\rho_{12}=2$}{13.7}{$\rho_{12}=81$}{0.916}{}
\DATASTATSYSREL{0.019\div 0.024}{3}{17900}{16.3}{$\rho_{13}=13$ $\rho_{23}=-17$}{24.0}{$\rho_{13}=35$ $\rho_{23}=29$}{0.882}{}
\hline
\hline
$y$&bin&\multicolumn{1}{c|}{$d\sigma/dy$}&
\multicolumn{1}{c|}{$\delta_{\text{stat}}$}&
\multicolumn{1}{c|}{$\rho_{\text{stat}}$}&
\multicolumn{1}{c|}{$\delta_{\text{syst}}$}&
\multicolumn{1}{c|}{$\rho_{\text{syst}}$}&
\multicolumn{1}{c|}{$1+\delta_{\text{hadr}}$} \\
 & &\multicolumn{1}{c|}{[pb]}&[$\%$]&\multicolumn{1}{c|}{[$\%$]}&
[$\%$]&\multicolumn{1}{c|}{[$\%$]}& \\
\hline
\DATASTATSYSREL{0.2\div 0.32}{1}{620}{16.4}{}{19.1}{}{0.858}{}
\DATASTATSYSREL{0.32\div 0.44}{2}{541}{15.9}{$\rho_{12}=-12$}{15.2}{$\rho_{12}=70$}{0.914}{}
\DATASTATSYSREL{0.44\div 0.56}{3}{408}{19.1}{$\rho_{13}=21$ $\rho_{23}=-42$}{18.6}{$\rho_{13}=37$ $\rho_{23}=70$}{0.957}{}
\DATASTATSYSREL{0.56\div 0.7}{4}{342}{18.0}{$\rho_{14}=3$ $\rho_{24}=33$ $\rho_{34}=-49$}{14.7}{$\rho_{14}=53$ $\rho_{24}=80$ $\rho_{34}=64$}{0.913}{}
\hline
\hline
$x_\gamma$&bin&\multicolumn{1}{c|}{$d\sigma/dx_\gamma$}&
\multicolumn{1}{c|}{$\delta_{\text{stat}}$}&
\multicolumn{1}{c|}{$\rho_{\text{stat}}$}&
\multicolumn{1}{c|}{$\delta_{\text{syst}}$}&
\multicolumn{1}{c|}{$\rho_{\text{syst}}$}&
\multicolumn{1}{c|}{$1+\delta_{\text{hadr}}$} \\
 & &\multicolumn{1}{c|}{[pb]}&[$\%$]&\multicolumn{1}{c|}{[$\%$]}&
[$\%$]&\multicolumn{1}{c|}{[$\%$]}& \\
\hline
\DATASTATSYSREL{0\div 0.3}{1}{65}{52.5}{}{55.5}{}{0.654}{}
\DATASTATSYSREL{0.3\div 0.6}{2}{180}{19.6}{$\rho_{12}=5$}{21.1}{$\rho_{12}=20$}{0.884}{}
\DATASTATSYSREL{0.6\div 0.8}{3}{397}{13.7}{$\rho_{13}=13$ $\rho_{23}=-2$}{18.8}{$\rho_{13}=-5$ $\rho_{23}=78$}{1.536}{}
\DATASTATSYSREL{0.8\div 1}{4}{367}{10.1}{$\rho_{14}=2$ $\rho_{24}=16$ $\rho_{34}=-21$}{13.5}{$\rho_{14}=-29$ $\rho_{24}=64$ $\rho_{34}=80$}{0.683}{}
\hline
\end{tabular}
\end{center}
\caption{Integrated diffractive dijet $ep$ cross section and bin averaged hadron level
  differential diffractive dijet $ep$ cross sections as a function of the
  variables $z_\pom$, $x_\pom$, $y$ and $x_\gamma$ for the 
  dijet photoproduction kinematic range in the phase space detailed in table
  \ref{tab_cuts}.
  For each data point, the statistical ($\delta_{\text{stat}}$) and
  systematic ($\delta_{\text{syst}}$) uncertainties, the
  corresponding correlation
  coefficients ($\rho_{\text{stat}}$, $\rho_{\text{syst}}$) and the
  hadronisation correction factors 
  ($1+\delta_{\text{hadr}}$) applied to the NLO calculations are given. 
 The overall normalisation uncertainty of $6\%$ is not included in the table.}
\label{tab_PhpXsec1}
\end{sidewaystable}

\begin{sidewaystable}[p]
\begin{center}
\footnotesize
%
%
%
%
%
\setlength{\tabcolsep}{0.3em}
\newcommand{\DATASTATSYSREL}[9]{$#1$&$#2$&$#3$&$#4$&#5&$#6$&#7&$#8$\\}
\begin{tabular}{|c|c|r|c|l|c|l|r|}
\hline
$E_T^{*\mathrm{jet1}}$&bin&\multicolumn{1}{c|}{$d\sigma/dE_T^{*\mathrm{jet1}}$}&
\multicolumn{1}{c|}{$\delta_{\text{stat}}$}&
\multicolumn{1}{c|}{$\rho_{\text{stat}}$}&
\multicolumn{1}{c|}{$\delta_{\text{syst}}$}&
\multicolumn{1}{c|}{$\rho_{\text{syst}}$}&
\multicolumn{1}{c|}{$1+\delta_{\text{hadr}}$} \\
{}[GeV]& &\multicolumn{1}{c|}{[pb/GeV]}&[$\%$]&\multicolumn{1}{c|}{[$\%$]}&
[$\%$]&\multicolumn{1}{c|}{[$\%$]}& \\
\hline
\DATASTATSYSREL{5.5\div 7}{1}{91}{14.9}{}{15.1}{}{0.877}{}
\DATASTATSYSREL{7\div 8.5}{2}{45.6}{21.1}{$\rho_{12}=-54$}{17.3}{$\rho_{12}=42$}{0.991}{}
\DATASTATSYSREL{8.5\div 10}{3}{11.2}{50.0}{$\rho_{13}=28$ $\rho_{23}=-69$}{25.3}{$\rho_{13}=76$ $\rho_{23}=25$}{0.956}{}
\DATASTATSYSREL{10\div 14.5}{4}{1.15}{63.8}{$\rho_{14}=-9$ $\rho_{24}=32$ $\rho_{34}=-63$}{66.4}{$\rho_{14}=44$ $\rho_{24}=90$ $\rho_{34}=28$}{0.840}{}
\hline
\hline
$M_X$&bin&\multicolumn{1}{c|}{$d\sigma/dM_X$}&
\multicolumn{1}{c|}{$\delta_{\text{stat}}$}&
\multicolumn{1}{c|}{$\rho_{\text{stat}}$}&
\multicolumn{1}{c|}{$\delta_{\text{syst}}$}&
\multicolumn{1}{c|}{$\rho_{\text{syst}}$}&
\multicolumn{1}{c|}{$1+\delta_{\text{hadr}}$} \\
{}[GeV]& &\multicolumn{1}{c|}{[pb/GeV]}&[$\%$]&\multicolumn{1}{c|}{[$\%$]}&
[$\%$]&\multicolumn{1}{c|}{[$\%$]}& \\
\hline
\DATASTATSYSREL{10\div 20}{1}{2.47}{32.3}{}{30.5}{}{0.899}{}
\DATASTATSYSREL{20\div 28}{2}{13.2}{12.7}{$\rho_{12}=-27$}{14.7}{$\rho_{12}=9$}{0.925}{}
\DATASTATSYSREL{28\div 36}{3}{12.4}{13.0}{$\rho_{13}=16$ $\rho_{23}=-28$}{18.0}{$\rho_{13}=-43$ $\rho_{23}=59$}{0.933}{}
\hline
\hline
$\vert\Delta\eta^{\text{jets}}\vert$&bin&\multicolumn{1}{c|}{$d\sigma/d\vert\Delta\eta^{\text{jets}}\vert$}&
\multicolumn{1}{c|}{$\delta_{\text{stat}}$}&
\multicolumn{1}{c|}{$\rho_{\text{stat}}$}&
\multicolumn{1}{c|}{$\delta_{\text{syst}}$}&
\multicolumn{1}{c|}{$\rho_{\text{syst}}$}&
\multicolumn{1}{c|}{$1+\delta_{\text{hadr}}$} \\
 & &\multicolumn{1}{c|}{[pb]}&[$\%$]&\multicolumn{1}{c|}{[$\%$]}&
[$\%$]&\multicolumn{1}{c|}{[$\%$]}& \\
\hline
\DATASTATSYSREL{0\div 0.5}{1}{171}{8.8}{}{12.1}{}{0.872}{}
\DATASTATSYSREL{0.5\div 1}{2}{147}{11.9}{$\rho_{12}=3$}{14.1}{$\rho_{12}=96$}{0.905}{}
\DATASTATSYSREL{1\div 1.5}{3}{93}{14.5}{$\rho_{13}=17$ $\rho_{23}=-5$}{18.4}{$\rho_{13}=85$ $\rho_{23}=94$}{0.936}{}
\DATASTATSYSREL{1.5\div 2}{4}{41}{25.5}{$\rho_{14}=13$ $\rho_{24}=18$ $\rho_{34}=2$}{29.4}{$\rho_{14}=79$ $\rho_{24}=89$ $\rho_{34}=95$}{0.978}{}
\hline
\hline
$\langle\eta^{\text{jets}}\rangle$&bin&\multicolumn{1}{c|}{$d\sigma/d\langle\eta^{\text{jets}}\rangle$}&
\multicolumn{1}{c|}{$\delta_{\text{stat}}$}&
\multicolumn{1}{c|}{$\rho_{\text{stat}}$}&
\multicolumn{1}{c|}{$\delta_{\text{syst}}$}&
\multicolumn{1}{c|}{$\rho_{\text{syst}}$}&
\multicolumn{1}{c|}{$1+\delta_{\text{hadr}}$} \\
 & &\multicolumn{1}{c|}{[pb]}&[$\%$]&\multicolumn{1}{c|}{[$\%$]}&
[$\%$]&\multicolumn{1}{c|}{[$\%$]}& \\
\hline
\DATASTATSYSREL{-1\div -0.45}{1}{48.4}{13.3}{}{13.0}{}{0.795}{}
\DATASTATSYSREL{-0.45\div -0.05}{2}{175}{10.1}{$\rho_{12}=1$}{12.8}{$\rho_{12}=81$}{0.881}{}
\DATASTATSYSREL{-0.05\div 0.25}{3}{264}{9.3}{$\rho_{13}=2$ $\rho_{23}=0$}{13.6}{$\rho_{13}=71$ $\rho_{23}=88$}{0.971}{}
\DATASTATSYSREL{0.25\div 0.65}{4}{134}{16.6}{$\rho_{14}=3$ $\rho_{24}=3$ $\rho_{34}=-6$}{22.3}{$\rho_{14}=46$ $\rho_{24}=79$ $\rho_{34}=90$}{0.976}{}
\hline
\end{tabular}
\end{center}
\caption{Bin averaged hadron level differential diffractive dijet $ep$ cross
  sections in the photoproduction kinematic range as a function of the
  variables $E_T^{*\text{jet1}}$, $M_X$, 
  $\vert\Delta\eta^{\text{jets}}\vert$ and
  $\langle\eta^{\text{jets}}\rangle$ in the phase space detailed 
  in table \ref{tab_cuts}.
  For each data point, the statistical ($\delta_{\text{stat}}$) and
  systematic ($\delta_{\text{syst}}$) uncertainties, the corresponding 
  coefficients ($\rho_{\text{stat}}$, $\rho_{\text{syst}}$) 
  and the hadronisation correction factors
  ($1+\delta_{\text{hadr}}$) applied to the NLO calculations are given. 
 The overall normalisation uncertainty of $6\%$ is not included in the table.}
\label{tab_PhpXsec2}
\end{sidewaystable}

\begin{sidewaystable}[p]
\begin{center}
\footnotesize
%
%
%
%
%
\setlength{\tabcolsep}{0.3em}
\newcommand{\DATASTATSYSREL}[9]{$#1$&$#2$&$#3$&$#4$&#5&$#6$&#7&$#8$\\}
\begin{tabular}{|c|c|r|c|l|c|l|r|}
\hline
$\vert\Delta\eta^{\text{jets}}\vert$&bin&\multicolumn{1}{c|}{$\sigma_{\gamma p}/\sigma_{DIS}$}&
\multicolumn{1}{c|}{$\delta_{\text{stat}}$}&
\multicolumn{1}{c|}{$\rho_{\text{stat}}$}&
\multicolumn{1}{c|}{$\delta_{\text{syst}}$}&
\multicolumn{1}{c|}{$\rho_{\text{syst}}$}&
\multicolumn{1}{c|}{$1+\delta_{\text{hadr}}$} \\
 & & &[$\%$]&\multicolumn{1}{c|}{[$\%$]}&
[$\%$]&\multicolumn{1}{c|}{[$\%$]}& \\
\hline
\DATASTATSYSREL{0\div 0.5}{1}{6.9}{14.9}{}{14.8}{}{0.867}{}
\DATASTATSYSREL{0.5\div 1}{2}{9.2}{21.3}{$\rho_{12}=-10$}{18.5}{$\rho_{12}=98$}{0.865}{}
\DATASTATSYSREL{1\div 1.5}{3}{7.1}{24.6}{$\rho_{13}=9$ $\rho_{23}=-12$}{22.7}{$\rho_{13}=95$ $\rho_{23}=97$}{0.908}{}
\DATASTATSYSREL{1.5\div 2}{4}{5.2}{38.5}{$\rho_{14}=5$ $\rho_{24}=10$ $\rho_{34}=-8$}{33.5}{$\rho_{14}=88$ $\rho_{24}=90$ $\rho_{34}=95$}{0.951}{}
\hline
\hline
$y$&bin&\multicolumn{1}{c|}{$\sigma_{\gamma p}/\sigma_{DIS}$}&
\multicolumn{1}{c|}{$\delta_{\text{stat}}$}&
\multicolumn{1}{c|}{$\rho_{\text{stat}}$}&
\multicolumn{1}{c|}{$\delta_{\text{syst}}$}&
\multicolumn{1}{c|}{$\rho_{\text{syst}}$}&
\multicolumn{1}{c|}{$1+\delta_{\text{hadr}}$} \\
 & & &[$\%$]&\multicolumn{1}{c|}{[$\%$]}&
[$\%$]&\multicolumn{1}{c|}{[$\%$]}& \\
\hline
\DATASTATSYSREL{0.2\div 0.32}{1}{8.0}{22.7}{}{22.5}{}{0.864}{}
\DATASTATSYSREL{0.32\div 0.44}{2}{7.7}{21.3}{$\rho_{12}=-12$}{18.4}{$\rho_{12}=87$}{0.912}{}
\DATASTATSYSREL{0.44\div 0.56}{3}{6.2}{24.1}{$\rho_{13}=12$ $\rho_{23}=-29$}{21.5}{$\rho_{13}=88$ $\rho_{23}=91$}{0.907}{}
\DATASTATSYSREL{0.56\div 0.7}{4}{8.9}{28.0}{$\rho_{14}=1$ $\rho_{24}=16$ $\rho_{34}=-33$}{14.7}{$\rho_{14}=84$ $\rho_{24}=84$ $\rho_{34}=75$}{0.841}{}
\hline
\hline
$z_\pom$&bin&\multicolumn{1}{c|}{$\sigma_{\gamma p}/\sigma_{DIS}$}&
\multicolumn{1}{c|}{$\delta_{\text{stat}}$}&
\multicolumn{1}{c|}{$\rho_{\text{stat}}$}&
\multicolumn{1}{c|}{$\delta_{\text{syst}}$}&
\multicolumn{1}{c|}{$\rho_{\text{syst}}$}&
\multicolumn{1}{c|}{$1+\delta_{\text{hadr}}$} \\
 & & &[$\%$]&\multicolumn{1}{c|}{[$\%$]}&
[$\%$]&\multicolumn{1}{c|}{[$\%$]}& \\
\hline
\DATASTATSYSREL{0\div 0.2}{1}{2.29}{42.7}{}{53.5}{}{0.697}{}
\DATASTATSYSREL{0.2\div 0.4}{2}{6.1}{20.4}{$\rho_{12}=-18$}{19.1}{$\rho_{12}=80$}{0.790}{}
\DATASTATSYSREL{0.4\div 0.6}{3}{8.5}{20.7}{$\rho_{13}=9$ $\rho_{23}=-26$}{16.3}{$\rho_{13}=85$ $\rho_{23}=92$}{0.932}{}
\DATASTATSYSREL{0.6\div 0.8}{4}{21.4}{43.0}{$\rho_{14}=0$ $\rho_{24}=8$ $\rho_{34}=-31$}{22.7}{$\rho_{14}=81$ $\rho_{24}=75$ $\rho_{34}=89$}{1.116}{}
\hline
\hline
$E_T^{*\mathrm{jet1}}$&bin&\multicolumn{1}{c|}{$\sigma_{\gamma p}/\sigma_{DIS}$}&
\multicolumn{1}{c|}{$\delta_{\text{stat}}$}&
\multicolumn{1}{c|}{$\rho_{\text{stat}}$}&
\multicolumn{1}{c|}{$\delta_{\text{syst}}$}&
\multicolumn{1}{c|}{$\rho_{\text{syst}}$}&
\multicolumn{1}{c|}{$1+\delta_{\text{hadr}}$} \\
{}[GeV] & & &[$\%$]&\multicolumn{1}{c|}{[$\%$]}&
[$\%$]&\multicolumn{1}{c|}{[$\%$]}& \\
\hline
\DATASTATSYSREL{5.5\div 7}{1}{8.0}{17.1}{}{20.4}{}{0.872}{}
\DATASTATSYSREL{7\div 8.5}{2}{8.05}{26.7}{$\rho_{12}=-39$}{12.2}{$\rho_{12}=51$}{0.957}{}
\DATASTATSYSREL{8.5\div 10}{3}{4.4}{61.6}{$\rho_{13}=13$ $\rho_{23}=-52$}{28.3}{$\rho_{13}=70$ $\rho_{23}=90$}{0.858}{}
\DATASTATSYSREL{10\div 14.5}{4}{2.4}{78.2}{$\rho_{14}=-4$ $\rho_{24}=16$ $\rho_{34}=-55$}{63.5}{$\rho_{14}=57$ $\rho_{24}=76$ $\rho_{34}=87$}{0.859}{}
\hline
\end{tabular}
\end{center}
\caption{Ratios of differential diffractive dijet $ep$ cross
  sections, measured in photoproduction, to measurements in DIS as a
  function of the variables $\vert\Delta\eta^{\text{jets}}\vert$, $y$,
  $z_\pom$ and $E_T^{*\text{jet1}}$ in the phase space detailed in
  table \ref{tab_cuts}.
  For each data point, the statistical ($\delta_{\text{stat}}$) and
  systematic ($\delta_{\text{syst}}$) uncertainties, the corresponding correlation
  coefficients ($\rho_{\text{stat}}$, $\rho_{\text{syst}}$) and the
  hadronisation correction factors 
  ($1+\delta_{\text{hadr}}$) applied to the NLO calculations are
  given.
}
\label{tab_xsecRatio}
\end{sidewaystable}

\FloatBarrier

\end{document}